\documentclass[twocolumn]{aastex61}
\submitjournal{ApJ}

\usepackage{hyperref}
\usepackage{amsmath}
\usepackage{amssymb}
\usepackage{graphicx}
\usepackage[utf8]{inputenc}

\newcommand{\thestar}{J0023$+$0307}

\usepackage[utf8]{inputenc}

\shorttitle{J0023+0307 -- A Second-Generation Main-Sequence Star with $\mbox{[Fe/H]}<-6$}
\shortauthors{Frebel et al.}
\begin{document}

\title{Chemical abundance Signature of J0023+0307 -- A Second-Generation \\Main-Sequence Star with $\mbox{[Fe/H]}<-6$   
  \footnote{This paper includes data gathered with the 6.5 meter
  Magellan Telescopes located at Las Campanas Observatory, Chile.}}                         

\newcommand{\alex}[1]{\textcolor{orange}{(APJ: #1)}}
\newcommand{\alexs}[2]{\textcolor{orange}{(APJ: \sout{#1} #2)}}

\correspondingauthor{Anna Frebel}
\email{afrebel@mit.edu}

\author{Anna Frebel}
\affiliation{Department of Physics \& Kavli Institute for Astrophysics and Space Research, Massachusetts Institute of Technology, Cambridge, MA 02139, USA}
\affiliation{Joint Institute for Nuclear Astrophysics - Center for Evolution of the Elements, East Lansing, MI 48824, USA}                        

\author{Alexander P. Ji}
\affiliation{The Observatories of the Carnegie Institution for Science, Pasadena, CA 91101, USA}
\affiliation{Hubble Fellow}

\author{Rana Ezzeddine}
\affiliation{Joint Institute for Nuclear Astrophysics - Center for Evolution of the Elements, East Lansing, MI 48824, USA}  
\affiliation{Department of Physics \& Kavli Institute for Astrophysics and Space Research, Massachusetts Institute of Technology, Cambridge, MA 02139, USA}

\author{Terese T. Hansen}
\affiliation{The Observatories of the Carnegie Institution for Science, Pasadena, CA 91101, USA}

\author{Anirudh Chiti}
\affiliation{Department of Physics \& Kavli Institute for Astrophysics and Space Research, Massachusetts Institute of Technology, Cambridge, MA 02139, USA}

\author{Ian B. Thompson}
\affiliation{The Observatories of the Carnegie Institution for Science, Pasadena, CA 91101, USA}

\author{Thibault Merle}
\affiliation{Institut d'Astronomie et d'Astrophysique, Universit\'{e} Libre de Bruxelles, CP 226, Boulevard du Triomphe, 1050 Brussels, Belgium}

\begin{abstract}
We present a chemical abundance analysis of the faint halo metal-poor main-sequence star J0023+0307, with $\mbox{[Fe/H]}<-6.3$, based on a high-resolution ($R\sim35,000$) Magellan/MIKE spectrum. The star was originally found to have $\mbox{[Fe/H]}< -6.6$ based on a Ca\,II\,K measurement in an $R\sim2,500$ spectrum. No iron lines could be detected in our MIKE spectrum. Spectral lines of Li, C, Na, Mg, Al, Si, and Ca were detected. The Li abundance is close to the Spite Plateau, $\log \epsilon$(Li) = 1.7, not unlike that of other metal-poor stars although in stark contrast to the extremely low value found e.g., in HE~1327$-$2326 at a similar [Fe/H] value. The carbon G-band is detected and indicates strong C-enhancement, as is typical for stars with low Fe abundances. Elements from Na through Si show a strong odd-even effect, and \thestar\ displays the second-lowest known [Ca/H] abundance. Overall, the abundance pattern of J0023+0307 suggests that it is a second-generation star that formed from gas enriched by a massive Population\,III first star exploding as a fall-back supernova 
The inferred dilution mass of the ejecta is $10^{5\pm 0.5}\,M_\odot$ of hydrogen, strongly suggesting J0023+0307 formed in a recollapsed minihalo. \thestar\ is likely very old because it has a very eccentric orbit with a pericenter in the Galactic bulge.
\end{abstract}


\section{Introduction}\label{intro}

The most metal-poor stars are tracers of the physical and chemical conditions of the early universe. In their atmospheres, they carry a record of this early time that was governed by the first stars, first supernovae, and the formation of the first galaxies \citep{Beers05,Frebel15}.
Much progress has been made in the past two decades to systematically uncover the extremely rare most metal-poor stars through systematic searches, and to chemically characterize them using large telescopes equipped with high-resolution spectrographs.
There are now about ${\sim}30$ stars known with $\mbox{[Fe/H]}\lesssim-4.0$ of which ${\sim}10$ have $\mbox{[Fe/H]}\lesssim-4.5$ \citep{HE0107_Nature, frebeletal05, he0557, caffau11, hansen14,keller14, Bonifacio15,allendeprieto15,frebel15b,aguado18b,Starkenburg18}.
Given their small Fe abundances, they are believed to be second-generation objects, i.e., only one progenitor was responsible for the elements now observed in each of these stars. 
This, in turn, enables studies of the first supernovae and the associated nucleosynthesis, by studying the stellar chemical abundance patterns and comparing them with theoretical predictions of the yields of these first explosions \citep{UmedaNomotoNature,iwamoto_science,heger10, placco15b,placco16}. 
This provides one of few means to gain clues on the nature of properties of the first stars. 
One striking feature besides the low Fe abundance is that nearly all of these stars are enhanced in carbon, suggesting that the first stars produce copious amounts of carbon, either during late stages of stellar evolution and/or during their explosions \citep{UmedaNomotoNature,meynet06,Choplin18}. 
Recent searches for these most iron-poor stars include those with the SkyMapper telescope \citep{keller14,Jacobson15}, the Pristine survey \citep{Starkenburg17}, the Best and Brightest survey \citep{Schlaufman14}, ToPoS \citep{Caffau13}, and based on SDSS data \citep{aguado16}. 

J0023+0307 was discovered and first observed by \citet{aguado18} with medium-resolution spectroscopy. A very weak Ca\,K line was detected (together with a significant contribution of interstellar Ca) but no Ca abundance was presented. Instead just an upper limit on the Fe abundance was listed, $\mbox{[Fe/H]}<-6.6$. No Fe lines were detected in their spectrum. Then, \citet{francois18} found $\mbox{[Ca/H]} = -5.7$, inferring $\mbox{[Fe/H]} < -6.1$ under the assumption $\mbox{[Ca/Fe]} = 0.4$ as no Fe lines were detected.
In this paper, we report new high-resolution spectroscopic observations of this star. 
Still, no Fe lines could be detected. The LTE limit (obtained from the strongest line at 3859\,{\AA}) is $\mbox{[Fe/H]}<-5.6$ (the NLTE limit is $\mbox{[Fe/H]}<-5.2$) and our Ca-abundance derived limit is $\mbox{[Fe/H]}<-6.3$. These results confirm that J0023+0307 is part of the group of stars with the lowest Fe abundances known.  

\section{Observations}

\begin{figure*}[!ht]
 \begin{center}
\includegraphics[clip=true,width=5.5cm,viewport=30 310 245 558]{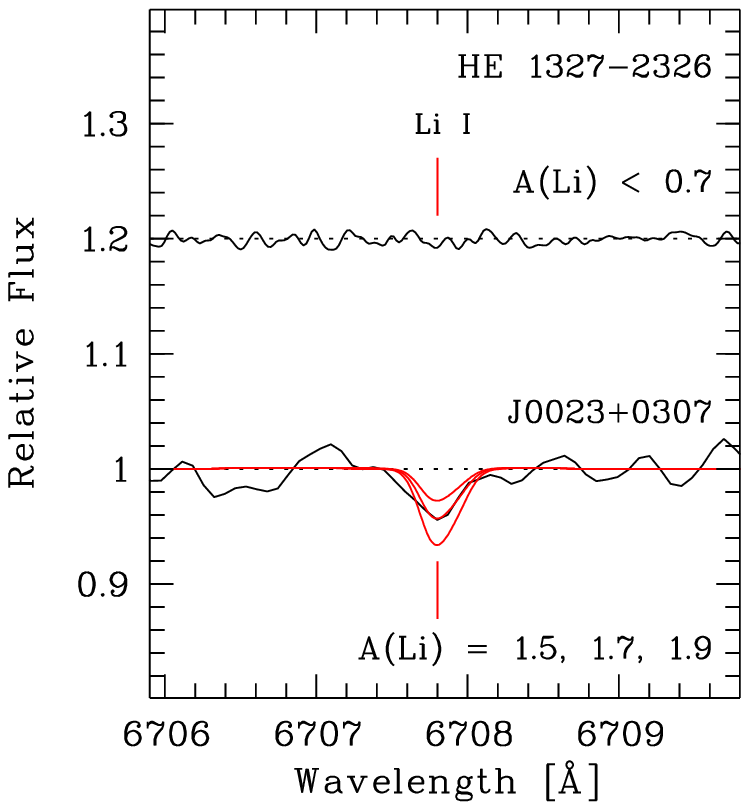}
\includegraphics[clip=true,width=5.5cm, viewport=30 393 245 628]{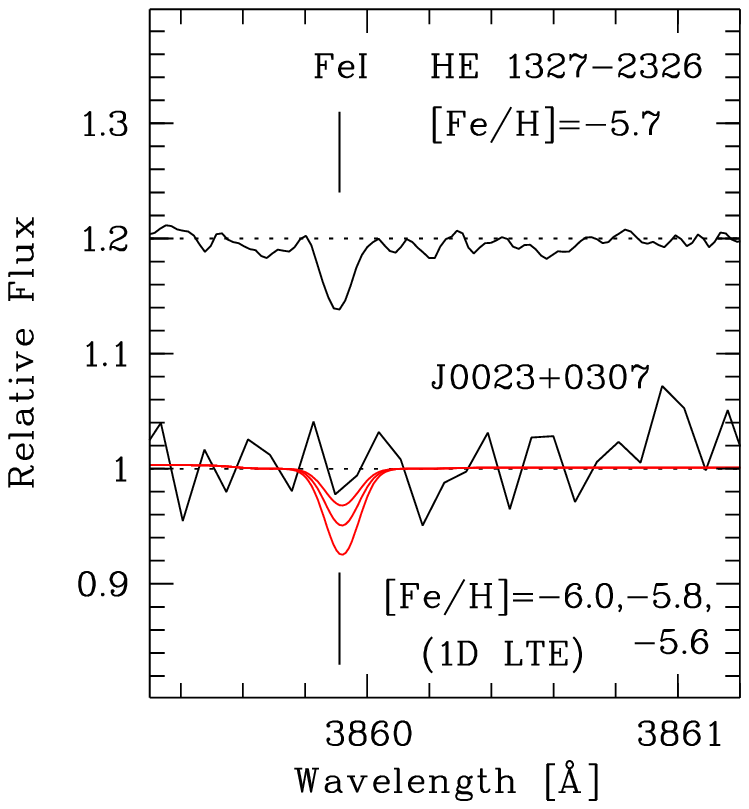}\\
  \includegraphics[clip=true,width=11.cm, viewport=30 133 500 378]{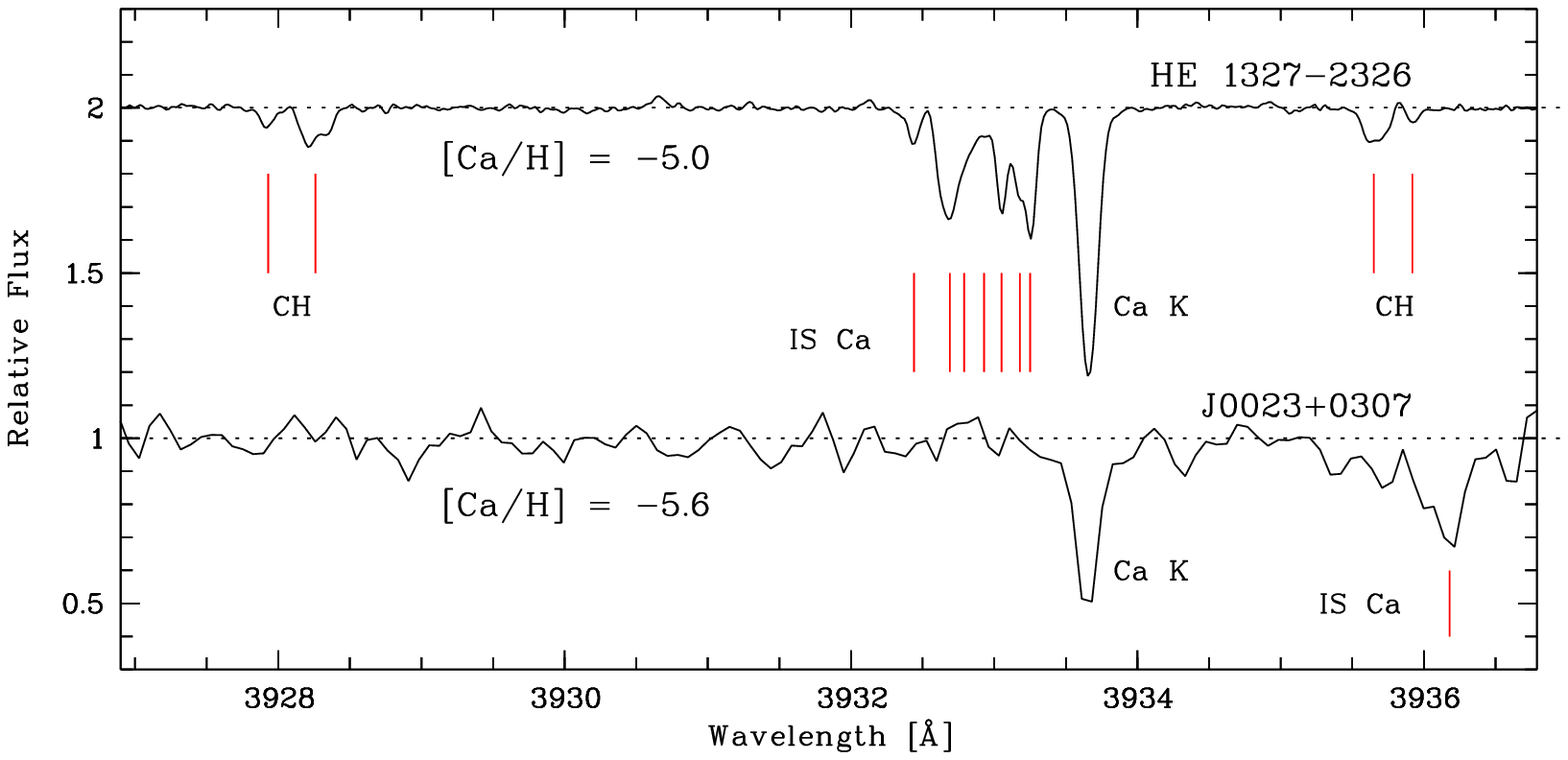}\\ 
\includegraphics[clip=true,width=11.1cm, viewport=30  396 500 628]{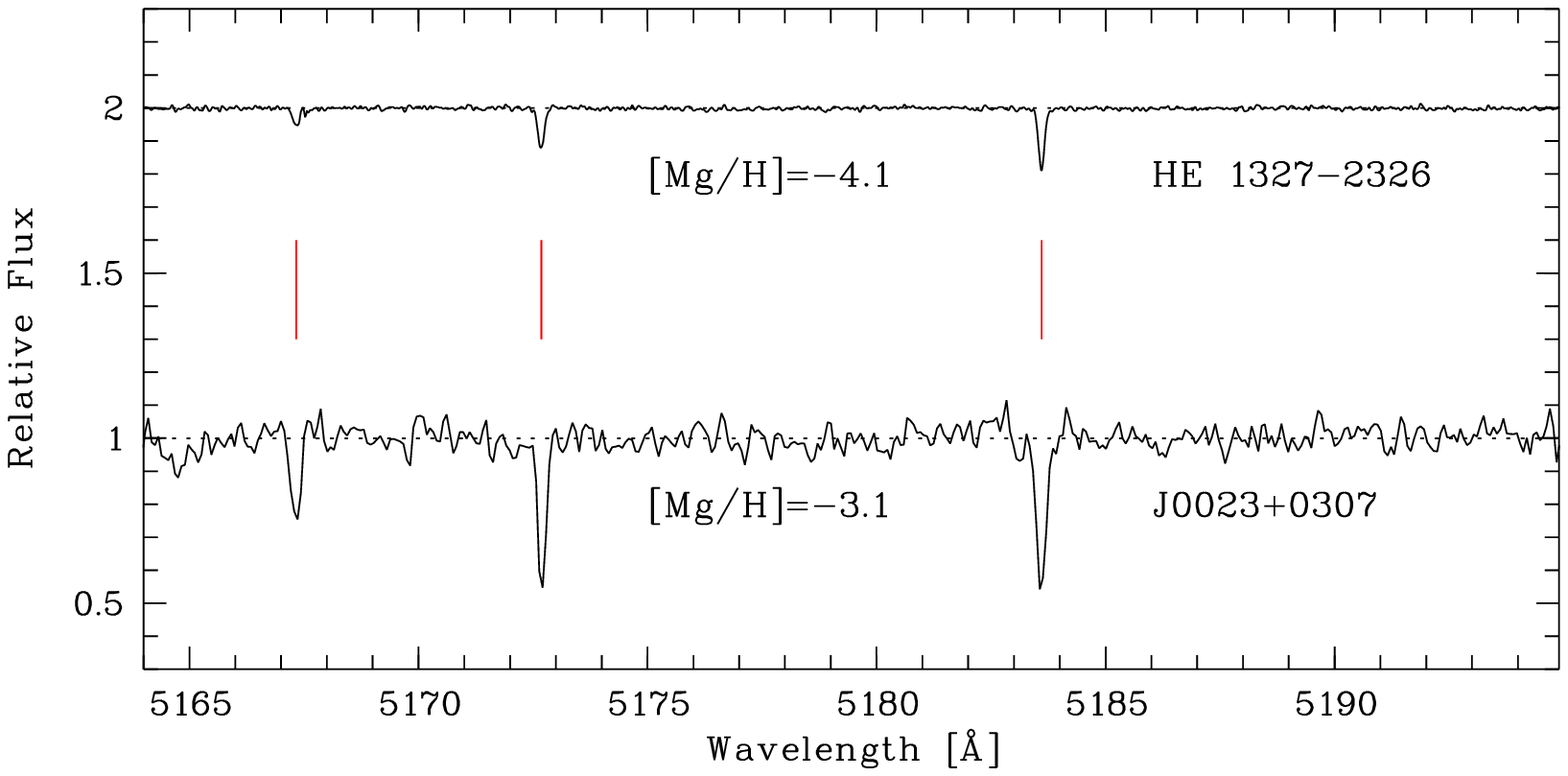}\\ 
\includegraphics[clip=true,width=15.1cm, viewport=10  384 513 590]{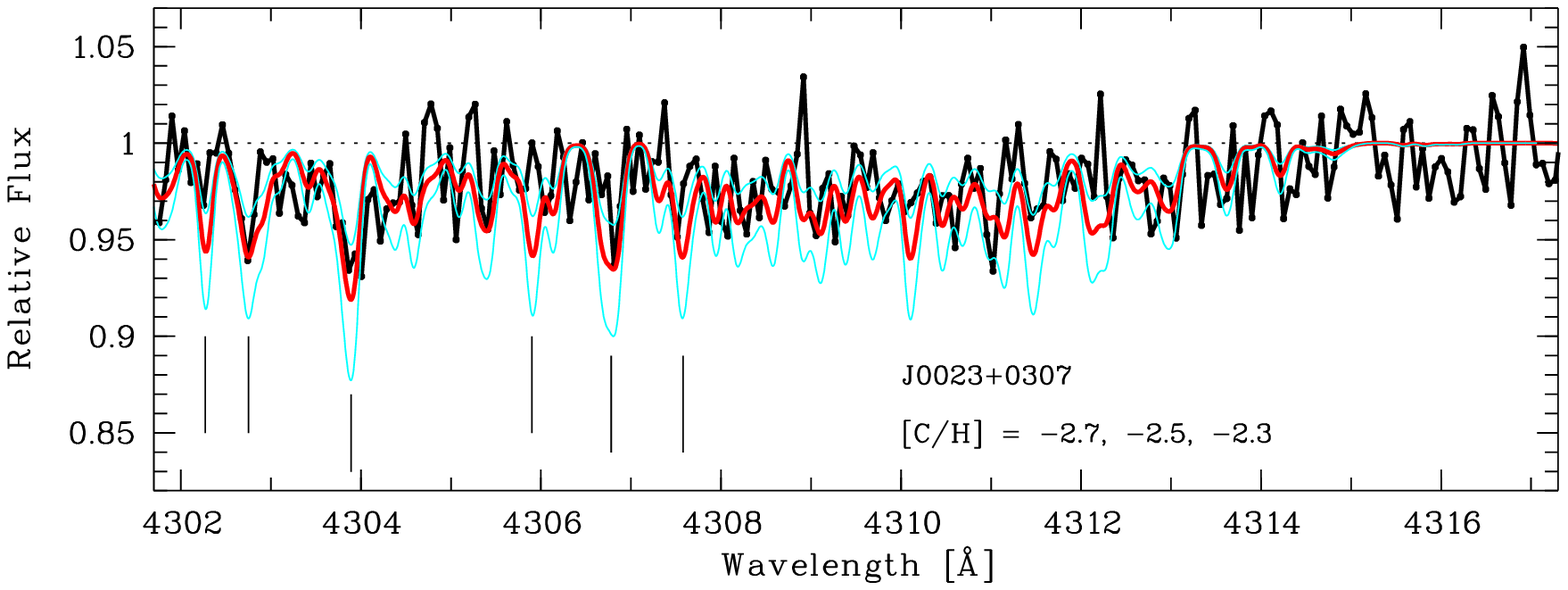}\\ 
\figcaption{\label{specs} Portions of the Magellan/MIKE spectrum of
    J0023+0307 in comparison with HE~1327$-$2326 (except for the CH region) near the Li\,I doublet at 6707\,{\AA} and the Fe\,I line at 3859\,{\AA} (top), the Ca\,II\,K line at 3933\,{\AA}, around the Mg\,b lines at 5170\,{\AA}, and the CH G-band at $\sim$4313 (bottom). Some absorption lines are indicated. See text for discussion.}
\end{center}
\end{figure*}

J0023+0307  (R.A. = 00:23:14.0 , Dec. = +03:07:58.1, $g=17.9$) was observed  with the MIKE spectrograph \citep{mike} on the Magellan-Clay telescope at Las Campanas Observatory on 2018 July 6-8, 15, and 24-25, and Aug 31-Sept 2, using a  $0\farcs7$ slit and $2\times2$ binning. The total exposure time was $\sim$24\,h but only $\sim20$\,h yielded useful data, i.e., with seeing of $\lesssim 1\farcs3$. Otherwise, the seeing was $\lesssim 0\farcs8$ for $\sim$11\,h, and  $\lesssim 1\farcs1$ for $\sim$9\,h. 
The spectral resolution is $\sim30,000$ in the red and $\sim35,000$ in the blue wavelength regimes. The spectrum covers 3050\,{\AA} to 9000\,{\AA} but is only usable above $\sim$3700\,{\AA}. Data were reduced with the CarPy pipeline \citep{Kelson03}\footnote{Available at http://obs.carnegiescience.edu/Code/python}. The resulting $S/N$ per pixel is 
  $\sim$40 at $\sim4000$\,{\AA}, 
  $\sim$60 at $\sim4700$\,{\AA}, 
  $\sim$35 at $\sim5200$\,{\AA}, and 
  $\sim$75 at $\sim6700$\,{\AA}. 
  We show portions of the spectrum around the Li\,I doublet at 6706\,{\AA}, the Fe\,I\,K line at 3859\,{\AA}, the Ca\,II\,K line at 3933\,{\AA}, the Mg\,b lines at 5170\,{\AA}, and G-band head region around 4313\,{\AA}  in Figure~\ref{specs}.
We also show the spectrum of HE 1327-2326 \citep{frebeletal05} for comparison purposes, which has a very similar effective temperature ($T_{\rm eff} = 6180-$).

  The heliocentric radial velocity obtained from the position of the Mg\,b lines is $-194.6 \pm 1.2$\,km\,s$^{-1}$.
  We see no evidence for velocity variations in our MIKE data over the course of two months.
  \citet{aguado18} quoted $v=-110 \pm 9$\,km\,s$^{-1}$ from the BOSS discovery spectrum, but this appears to be due to a missed conversion between vacuum and air wavelengths.
  BOSS reports a velocity of $-180 \pm 7$\,km\,s$^{-1}$ \citep{dawson13}, but this velocity relies on broad Balmer lines, uses an inappropriate template (a metal-rich B star), and also may be affected by ISM absorption lines. Our analysis of several H lines in the BOSS spectrum yields a velocity of $-205 \pm 10$\,km\,s$^{-1}$, consistent with the MIKE velocity.
  There are also two X-Shooter \citep{Vernet11} observations available on the VLT archive\footnote{Based on observations collected at the European Southern Observatory under ESO programme(s) 099.D-0576(A).} \citep{francois18}. We independently reduced these data with the default X-Shooter pipeline \citep{Goldoni06}. The two spectra were observed 2017 Jun 30 and 2017 Jul 21, with velocities $-198.7$ and $-192.5$\,km\,s$^{-1}$, showing no significant deviation from our MIKE velocities.

\section{Stellar Parameters}

{\thestar} has no detected Ti, Fe, or Ca\,I lines. It is thus not possible to determine stellar parameters through standard spectroscopic methods.
We determined an effective temperature from SDSS photometry \citep{ahn12}. {\thestar} has $g,\,r,\,i,\, = 17.90 \pm 0.01, 17.62 \pm 0.01, 17.52 \pm 0.01$.
The dust maps from \citet{Schlafly11} give reddening values $A(g)=0.108$, $A(r)=0.075$, $A(i)=0.056$, with a color uncertainty of $0.02$ mag.
We convert the dereddened magnitudes to $B-V$ and $V-I$ colors with the conversions in \citet{jordi06}, then apply the color-temperature relations from \citet{casagrande10} assuming $\mbox{[Fe/H]} = -5$.
For $B-V$ and $V-I$, we obtain $T_{\rm eff} = 6260 \pm 140\,$K and $T_{\rm eff} = 5997 \pm 130\,$K, respectively.
The reddening uncertainty dominates the temperature uncertainty.
Since the $B-V$ color-temperature relation is extremely sensitive to metallicity, and the color-temperature relations are not well-calibrated at the lowest metallicity \citep{casagrande10}, we adopt the temperature from the $V-I$ color which has minimal dependence on metallicity (${\approx}\,50$\,K from $\mbox{[Fe/H]}=-5$ to $\mbox{[Fe/H]}=-3$).

We determined $\log g$ with the equation \citep{venn17}:
\begin{align*}
\log g = 4.44 &+ \log M_\star + 4 \log (T_{\rm eff}/5780) \\ &+ 0.4 (g_0 - \mu + BC(g) - 4.75)
\end{align*}
We use $g_0 = 17.79$ and $T_{\rm eff} = 5997 \pm 130\,$K from above.
We adopt $M_\star = 0.5-0.6\,M_\odot$ as a reasonable range for a 12\,Gyr old main-sequence turnoff star.
For the distance modulus, we use the parallax-based distance from {\it Gaia} DR2 inferred by \citet{BailerJones18}, giving $\mu = 11.68^{+0.54}_{-0.49}$ \citep{GaiaSatellite,GaiaDR2}.
The bolometric correction for the SDSS $g$ band from \citet{Casagrande14} is $BC(g) = -0.33 \pm 0.03$, where the error covers differences from varying the temperature within its error bars.
In total, this gives $\log g = 4.61-4.69 \pm 0.25$, where the range is changing the mass from $0.5-0.6\,M_\odot$ and the uncertainty is dominated by the parallax uncertainty. Both of these values suggest {\thestar} to be a main-sequence star. In the following we thus adopt $\log g=4.6$, taking into account the unevolved nature and hence a lower mass for the star. The surface gravity has little influence on any of the derived chemical abundances for such warm stars, so this choice is not critical. For the microturbulence, we adopt $v_{\rm mic} = 1.5$, a value typical for warm stars \citep{barklem05, aoki13}. Given that only weak lines are available and that the star is fairly warm, this choice hardly affects any of the determined abundances.
For comparison, \citet{aguado18} adopted $T_{\rm eff} = 6188 \pm 84$\,K, $\log g=4.9\pm 0.5$ and $v_{\rm mic} = 2.0$ from their analysis of a $R \sim 2,500$ ISIS spectrum.

\section{The Chemical Abundance Pattern of J0023+0307 -- iron-poor but not as metal-poor}\label{abund}

\subsection{LTE Abundance determination}

To calculate abundances we used the 2017 version of \texttt{MOOG} including Rayleigh scattering \citep{Sneden73, Sobeck11}, and using Barklem damping coefficients
\footnote{\href{url}{https://github.com/alexji/moog17scat}}.
Our analysis software (SMH; \citealt{casey14}) employs a 1D plane-parallel model atmosphere with $\alpha$-enhancement \citep{Castelli04} and assumes local thermal equilibrium (LTE). Equivalent widths were measured for lines that were detected (Na, Mg, Al, Si, Ca) and abundances were calculated correspondingly. For Li, C, and upper limits of other elements (e.g., Ti, Fe, Sr, Ba), a spectrum synthesis approach was chosen. Line abundances are listed in Table~\ref{Tab:eqw}. Lower limits on  [X/Fe] abundances were calculated using solar abundances from \citet{Asplund09}, and are given in Table~\ref{Tab:abund}. 
In the following, we comment on selected elements and their abundance determinations.

We tentatively detect Li with an abundance A(Li)$=1.7$. In Figure~\ref{specs}, we show the Li line and synthetic spectra for A(Li)$=1.7 \pm 0.2$. There is clearly a significant feature at the position of the Li doublet at 6707\,{\AA}, but the line profile is somewhat distorted, and the overall depth is only ${\approx}3\%$. Several other apparent absorption features of similar depth occur near the $\lambda$ 6707 line, but all of these are clearly associated with imperfect sky subtraction residuals. Indeed, the sky spectrum shows no features at the Li line position which suggests that the line is real. Still, we caution that the Li measurement is rather uncertain and deserves further study based on additional data.

\begin{deluxetable}{llrrrrc}
\tablewidth{0pt}
\tablecaption{\label{Tab:eqw} Equivalent Widths Measurements}
\tablehead{
\colhead{$\lambda$} & \colhead{Species} & \colhead{$\chi$} & \colhead{$\log{gf}$} & \colhead{EW} & \colhead{$\log\epsilon(\rm X)$} & \colhead{$\log\epsilon(\rm X)$} \\
\colhead{[{\AA}]} &  \colhead{} & \colhead{[eV]} & \colhead{[dex]} &\colhead{[m{\AA}]} & \colhead{[dex LTE]} & \colhead{[dex NLTE]}}
\startdata
 6707    &    Li\,I &   \nodata &  \nodata &    17  & 1.70\tablenotemark{a}    & \nodata \\
 4313    &    CH    &   \nodata &  \nodata &   syn  & 5.89 & \nodata \\
 5889.95 & Na\,I\tablenotemark{a}& 0.00 &   0.11   &  19.5: & 2.26:  & 2.32:  \\
 5895.92 & Na\,I\tablenotemark{a}& 0.00 &$-$0.19   &  10.0: & 2.23:  & 2.31:  \\
 3829.36 &    Mg\,I &      2.71 &  $-$0.21 &   105.4& 4.52 & 4.64\\
 3832.30 &    Mg\,I &      2.71 &     0.27 &   168.5& 4.62 & 4.81\\
 3838.29 &    Mg\,I &      2.72 &     0.49 &   207.9& 4.65 & 4.78\\
 4702.99 &    Mg\,I &      4.34 &  $-$0.38  &   16.5&  4.74 &  4.98   \\
 5167.32 &    Mg\,I &      2.71 &  $-$1.03 &    65.1& 4.68 & 4.82 \\
 5172.68 &    Mg\,I &      2.71 &  $-$0.45 &    98.1& 4.49 & 4.60 \\
 5183.60 &    Mg\,I &      2.72 &  $-$0.24 &   128.6& 4.60 & 4.65 \\
 5528.41 &    Mg\,I &      4.34 &  $-$0.50 &    12.9& 4.73  & 4.95   \\
 3944.06 &    Al\,I &      0.00 &  $-$0.63 &    29.3& 2.40 & 2.80 \\
 3961.52 &    Al\,I &      0.01 &  $-$0.34 &    27.4& 2.21 & 2.97 \\
 3905.52 &    Si\,I &      1.91 &  $-$1.09 &    51.1& 3.94 & 4.49 \\
 3933.66 &   Ca\,II &      0.00 &     0.11 &    98.0& 0.57 & 0.62 \\\hline
 4226.73 &    Ca\,I &      0.00 &     0.24 &  $<$12    & $<$1.4  & \nodata \\
 4246.82 &    Sc\,II&      0.32 &     0.24 &  $<$8     & $<-$0.4 & \nodata \\
 3761.32 &   Ti\,II &      0.57 &     0.18 &  $<$12    & $<$0.6  & \nodata \\
 4254.33 &   Cr\,I  &      0.00 &  $-$0.11 &  $<$8     & $<$1.4  & \nodata \\
 4030.75 &    Mn\,I &      0.00 &  $-$0.48 &  $<$10    & $<$1.5  & \nodata \\
 3859.11 &    Fe\,I &      0.00 &  $-$0.71 &  $<$10    & $<$1.9  & $<$2.3  \\
 3858.30 &    Ni\,I &      0.42 &  $-$0.95 &  $<$10    & $<$2.2  & \nodata \\
 4077.71 &   Sr\,II &      0.00 &  0.15    &  $<$10    & $<-$1.5 & \nodata \\
 4554.03 &   Ba\,II &      0.00 &  0.16    &  $<$8     & $<-$1.3 & \nodata \\
\enddata
\tablenotetext{a}{See text for discussion.} 
\end{deluxetable}

For C, we closely investigated the CH G-band region between 4300 and 4325\,{\AA}. The S/N is moderate but several of the six strongest, most isolated CH features in this region (4302.27, 4302.75, 4303.89, 4305.90, 4306.78, and 4307.58\,{\AA}) are well detected. Other regions (4310-4313\,{\AA}) are consistent with the abundance from these strong lines. The observed spectrum with synthetic spectra (using the linelist of \citealt{masseron14}) overlaid is show in Figure~\ref{specs} (bottom panel). The resulting best fit abundance is $\mbox{[C/H]} = -2.5$. The CH feature at 4323\,{\AA} can be fit with the same abundance. We also tested to what extent the C abundance depends on the O abundance. We find no change in carbon abundance up to an assumed $\mbox{[O/H]}=-1.0$. For higher O values, small changes (0.05\,dex) begin to appear that would reduce the C abundance but such a high O abundance is not likely to be correct.
Hence, J0023+0307 appears to be another carbon-enhanced hyper-metal-poor star.
Our measurement is also in agreement with that of \citet{aguado18}, whose upper limit is $\mbox{[C/H]} < -2.3$. \citet{francois18} report a higher value of $\mbox{[C/H]} = -2.0$.

Both Na\,D lines are detected but their weak lines present as unusually broad. We thus consider the abundances as uncertain, even though both lines yield nearly identical abundances. We note that at least three, likely four prominent interstellar Na components are found, but they are well separated from the stellar lines. The interstellar features for Na and Ca\,II\,K have essentially identical shapes. While it is possible that an additional interstellar feature could blend with the stellar Na lines, it appears unlikely given that such gas would need to have the same high velocity as \thestar\ ($-195$\,km\,s$^{-1}$). 
Higher S/N data might thus yield a more accurate NA abundance.

\citet{aguado18} indirectly suggest that their [Ca/H] measurement would be [Ca/H] = [Fe/H] + [Ca/Fe] = $-6.2$. We derive $\mbox{[Ca/H]}=-5.8$, in agreement when taking into account that their value was derived from a spectrum with much lower resolution that could have been affected by the subtraction of an interstellar Ca component. The interstellar component is well-separated from the stellar Ca feature in our high-resolution spectrum, see Figure~1. \citet{francois18} report $\mbox{[Ca/H]}=-5.7$, in good agreement with out measurement.

\begin{figure}[!ht]
 \begin{center}
\includegraphics[clip=true,width=8.5cm,viewport=10 100 440 373]{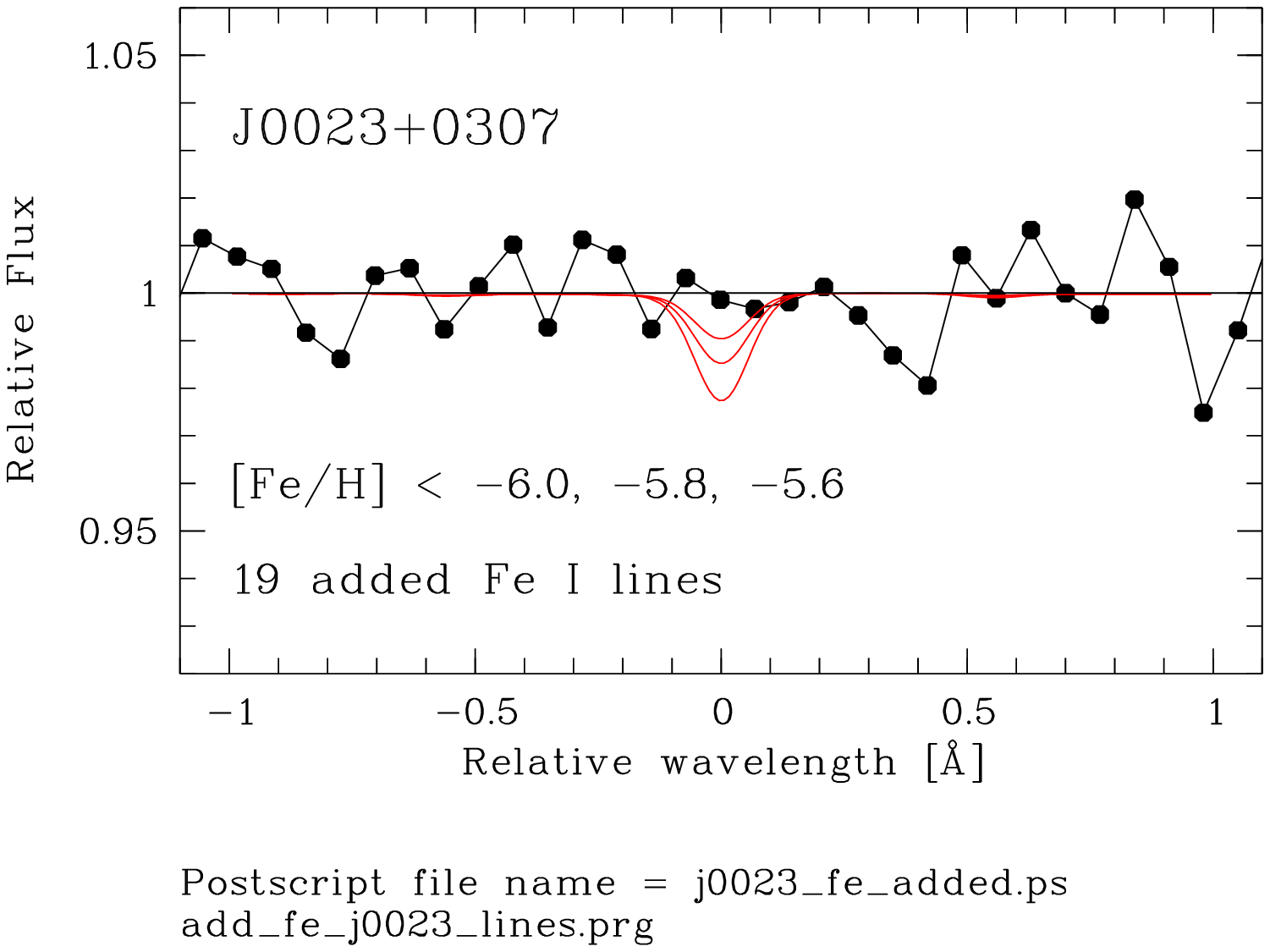}\\
\includegraphics[clip=true,width=8.5cm,viewport=10 100 440 373]{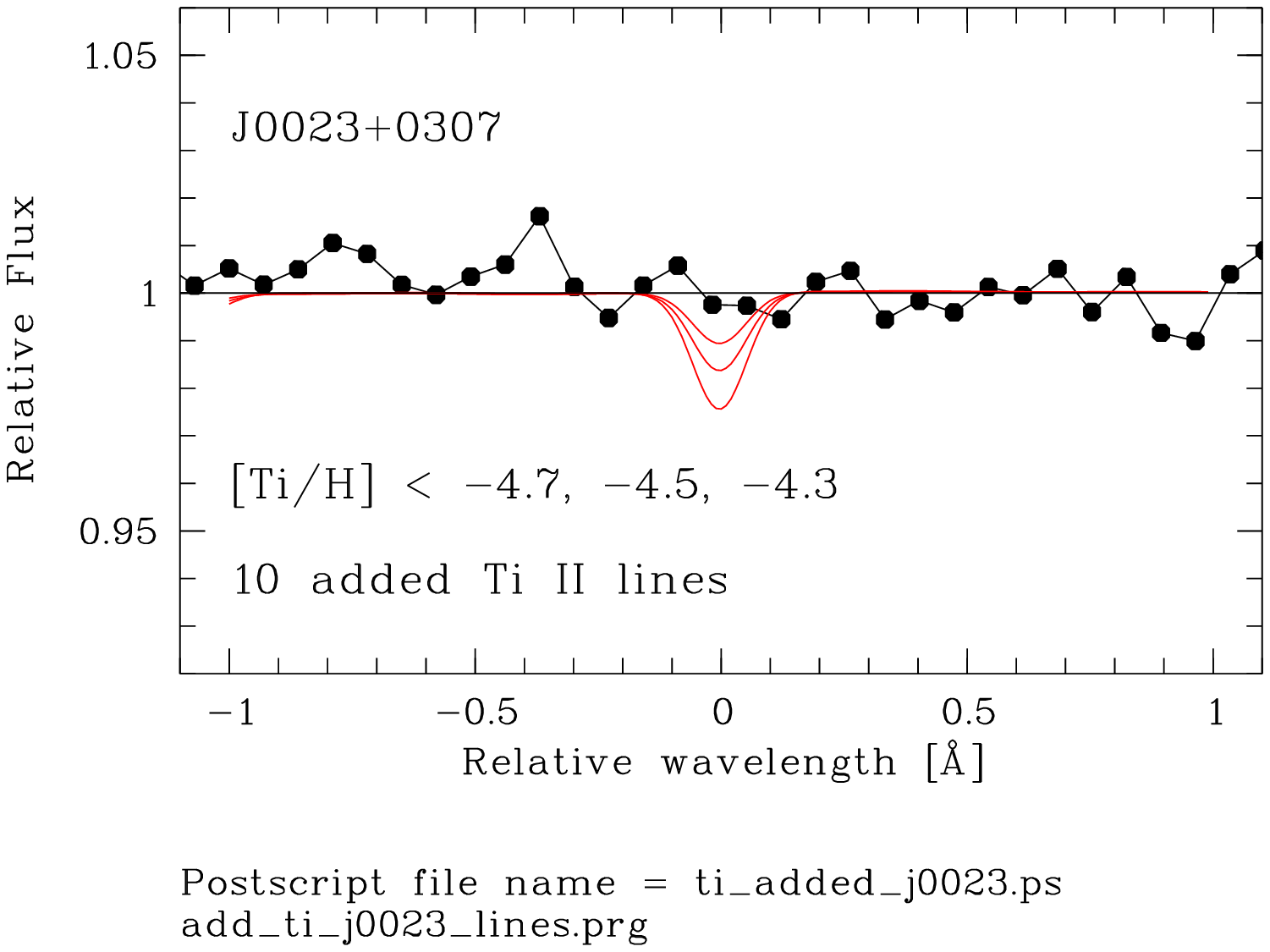}
\figcaption{\label{composite} Top: Composite spectrum of 19 Fe\,I lines overlaid with composite synthetic spectra of varying abundances as illustration.  
Bottom: Composite spectrum of 10 Ti\,II lines overlaid with composite synthetic  spectra of varying abundances. }
\end{center}
\end{figure}

Since no Fe lines were detected, we stacked portions of the spectrum that each contain a strong isolated Fe\,I line to investigate whether a signal would emerge in the composite spectrum. We followed the procedure outlined in \citet{frebel_he1300}. The same was done using synthetic spectra of varying abundances. We verified this procedure by stacking the spectral regions of four detected Mg\,I lines to reproduce the Mg abundance obtained from individual lines. Stacking 19 Fe lines between 3750 and 4050\,{\AA} did not yield a signal. Adding more lines from noisier, bluer parts of the spectrum did not change the outcome. By matching the noise level in the composite spectrum with the corresponding composite synthetic spectrum we derive an upper limits of $\mbox{[Fe/H]} < -5.8$. This is shown in Figure~\ref{composite}. We also stacked portions of the spectrum containing 10 Ti\,II lines. No signal was produced and the upper limit is $\mbox{[Ti/H]} < -4.5$.

\begin{deluxetable*}{lrrrrrcrrrr}
\tablewidth{0pt}
\tablecaption{\label{Tab:abund} Chemical abundances of {\thestar} and HE~1327$-$2326}
\tablehead{ 
\colhead{} & 
\multicolumn{5}{c}{{\thestar}} & 
\colhead{} & 
\multicolumn{4}{c}{HE~1327$-$2326} \\
\cline{2-6} \cline{8-11} 
\colhead{Species} & 
\colhead{[X/H]} & 
\colhead{[X/Fe]} & 
\colhead{[X/H]} & 
\colhead{[X/Fe]}& 
\colhead{$\sigma$} & 
\colhead{} & 
\colhead{[X/H]} & 
\colhead{[X/Fe]} &
\colhead{[X/H]} & 
\colhead{[X/Fe]}  \\
 & \multicolumn{2}{c}{(LTE)} & \multicolumn{2}{c}{(NLTE)} & & \multicolumn{2}{c}{(LTE)} & \multicolumn{2}{c}{(NLTE)}
}
\startdata
Li\,I  &\nodata&1.70\tablenotemark{a}& \nodata & \nodata  & 0.20 && \nodata & $<$0.70\tablenotemark{a} & \nodata & \nodata    \\
 CH    & $-$2.54   & $>$3.76  &\nodata & \nodata & 0.20 && $-$1.53 & 4.18 &\nodata & \nodata   \\
Na\,I  & $-$4.00:  & $>$2.30  & $-$3.92:&$>$1.98 & 0.30 && $-$3.25 & 2.46 & $-$3.30 & 1.90 \\
Mg\,I  & $-$2.97   & $>$3.33  & $-$2.82 &$>$3.08 & 0.10 && $-$4.06 & 1.65 & $-$3.80 & 1.40    \\
Al\,I  & $-$4.15   & $>$2.15  & $-$3.56 &$>$2.34 & 0.20 && $-$4.55 & 1.16 & $-$4.02 & 1.18    \\
Si\,I  & $-$3.55   & $>$2.75  & $-$3.02 &$>$2.88 & 0.20 && $-$4.50 & 0.70 & $-$4.31 & 0.89   \\
Ca\,II & $-$5.77   & $>$0.53  & $-$5.72 &$>$0.18 & 0.20 && $-$5.00 & 0.71 &\nodata &\nodata\\\hline
Sc\,II & $<-$3.5  & \nodata & \nodata &\nodata &\nodata && $<-$1.68 &$<$0.98 &\nodata & \nodata     \\
Ti\,II & $<-$4.3\tablenotemark{b}& \nodata & \nodata &\nodata &\nodata && $-$5.04 & 0.67 &\nodata & \nodata     \\
Ti\,II & $<-$4.5\tablenotemark{c}& \nodata & \nodata &\nodata &\nodata &&\nodata&\nodata&\nodata & \nodata     \\
Cr\,I  & $<-$4.2  & \nodata & \nodata &\nodata &\nodata&&$<-$5.19 &$<$0.52 & \nodata & \nodata\\
Mn\,I  & $<-$4.0  & \nodata & \nodata &\nodata &\nodata&&$<-$4.59 &$<$1.12 & \nodata & \nodata \\
Fe\,I  & $<-$5.6\tablenotemark{b}&\nodata & $<-$5.2 & \nodata &\nodata&& $-$5.71 &\nodata & $-$5.2\tablenotemark{d} & \nodata \\
Fe\,I  & $<-$5.8\tablenotemark{c}&\nodata &$<-$5.4&\nodata &\nodata&&\nodata &\nodata &\nodata & \nodata\\
Fe\,I  & $<-$6.3
\tablenotemark{e}&\nodata &$<-$5.9&\nodata &\nodata&&\nodata &\nodata &\nodata & \nodata\\
Ni\,I  & $<-$4.0  & \nodata &\nodata&\nodata &\nodata&& $-$5.49 & 0.22 & \nodata & \nodata\\
Sr\,II & $<-$4.3  & \nodata &\nodata&\nodata &\nodata&& $-$4.63 & 1.08 & \nodata & \nodata \\
Ba\,II & $<-$3.5  & \nodata &\nodata&\nodata &\nodata&& $-$4.32 & 1.39 & \nodata & \nodata \\
\enddata
\tablecomments{HE~1327$-$2326 abundances taken from \citet{collet06,frebel08}, except for Si from \citet{ezzeddine18}.}
\tablenotetext{a}{A(Li) is given instead of [X/Fe]. See text for discussion.}
\tablenotetext{b}{Obtained from the Ti\,II line at 3761\,{\AA}, and Fe\,I line at 3859\,{\AA}, respectively.}
\tablenotetext{c}{Obtained from the composite spectrum of eight Ti\,II lines, and 16 Fe\,I lines, respectively.}
\tablenotetext{d}{Adopted from \citet{ezzeddine18}.}
\tablenotetext{e}{Final adopted Fe abundance. See text for discussion. This value is used to calculate [X/Fe].}
\end{deluxetable*}

Considering that the other three stars with $\mbox{[Fe/H]} < -5.0$ (based on detected Fe lines) all have Ca abundances in excess of iron, we use this as an additional constraint on the iron abundance of \thestar. HE~1327$-$2326 has $\mbox{[Ca/Fe]} = 0.71$ \citep{frebel08}, HE~0107$-$5240 has $\mbox{[Ca/Fe]} = 0.46$ \citep{christliebetal04}, and SMSS0313$-$6707 has $\mbox{[Ca/Fe]} > 0.34$ \citep{bessell15} (considering 1D LTE values). We thus assume a ratio of $\mbox{[Ca/Fe]} \gtrsim 0.5$ for \thestar, given the many chemical similarities in these four stars compared to those with iron abundances of $\mbox{[Fe/H]} \gtrsim -5.0$ which begin to show more regular halo abundance patterns. We thus adopt $\mbox{[Fe/H]} < -6.3$ as our final upper limit for the iron abundance. This places \thestar\ firmly among the most iron-poor stars yet discovered. Future additional data on \thestar\ should confirm this value, or even push it to lower values, as e.g., in the case of SMS0313$-$6707. Following the same argument, \citet{Bonifacio15} detected no Fe lines in the spectrum of star SDSS~J1035$+$0641 but scaling the very low measured Ca abundance suggests an Fe upper limit of $\mbox{[Fe/H]} < -5.6$ (consistent with their $1\sigma$ Fe limit).

We also determined additional upper limits for Sc, Cr, Mn, Ni, Sr and Ba, if values were below $\mbox{[X/H]}<-3.5$. They are given in Tables~\ref{Tab:eqw} and \ref{Tab:abund}.  Abundance uncertainties are also listed in Table~\ref{Tab:abund}. 
Mg is very well determined from eight lines, so its uncertainty is significantly smaller than for the other abundances derived from just one line or feature.


\subsection{NLTE abundances}
We also calculated Non-LTE (NLTE) abundances for {\thestar} (and effects of using NLTE over LTE upper limits) for relevant lines of the elements Na, Mg, Al, Si, Ca, and Fe. 

\subsubsection{NLTE methods}
NLTE abundances were computed for Na, Mg, Al, Si, Ca and Fe using the radiative transfer code \texttt{MULTI} in its 2.3 version \citep{Carlsson1986,carlsson1992} and \texttt{MARCS} model atmospheres \citep{gustafsson1975,gustafsson2008} interpolated to the adopted stellar parameters. 
Atomic models used in the NLTE calculations for Na, Al, Mg, Si and Ca were built uniformly using the code \texttt{Formato2.0}\footnote{https://github.com/thibaultmerle/formato} (T. Merle et al. in prep). We provide below a brief description of the atomic data used in the atoms.

\subsubsection{Model atoms}
The atomic models for Na, Mg, Al, Si and Ca used in our NLTE calculations include up-to-date atomic data from various databases. Energy levels and radiative bound-bound ($bb$) transitions for the neutral and first ionized species for each element (except for Ca\,II where only the first ionized species levels were included ) were extracted from the NIST\footnote{https://www.nist.gov/pml/atomic-spectra-database},  VALD3\footnote{http://vald.astro.uu.se} and Kurucz\footnote{http://kurucz.harvard.edu/linelists.html} atomic databases. The number of energy levels and radiative $bb$ lines used in each atom are found in Table\,\ref{Tab:nlte_atoms}.
R-matrix radiative photoionization transitions were also included for all the transitions, when available, from the TOPBASE\footnote{http://cdsweb.u-strasbg.fr/topbase/topbase.html} database.

All levels in our atoms are also coupled via inelastic collisional transitions by electrons and neutral hydrogen atoms. For collisional rates by electrons, we used the empirical impact parameter approximation by \citet{seaton1962a}. For  inelastic hydrogen collisions, which can have important effects on NLTE abundance calculations and possibly produce large uncertainties if not properly treated \citep{barklem2010}, we used available rate coefficients computed via ab-initio quantum calculations for all our atoms, except for Ca\,II where the classical \citet{drawin1969a} equation was used. The rates for Ca\,II were scaled with a scaling factor S$_{\mathrm{H}}=0.1$ following \citet{mashonkina2013}. Regardless, the Ca\,II\,K line at 3933\,{\AA} used here is only little affected by NLTE \citep{sitnova2018}. The references for the hydrogen collisional data used for each element species are found in Table\,\ref{Tab:nlte_atoms}. For Fe, we used a comprehensive Fe\,I/Fe\,II model atom described in details in \citet{Ezzeddine2016}, with hydrogen collision rates incorporated from quantum calculations by \citet{barklem2018}.

\begin{deluxetable}{lrrl}
\tablewidth{0pt}
\tablecaption{\label{Tab:nlte_atoms} Model atoms used in the NLTE calculations}
\tablehead{
\colhead{Species} & \colhead{N$_{\mathrm{levels}}$} & \colhead{N$_{\mathrm{rad}}^{\mathrm{bb}}$} & \colhead{H collision rates ref.}}
\startdata
Na\,I & 139 & 443 & \citet{barklem2010}\\
Mg\,I\ & 229 & 1629 & \citet{belyaev2012} \\
       &     &      & and \citet{guitou2015}\\
Al\,I & 136 & 223 & \citet{belyaev2013}\\
Si\,I & 296 & 9503 & \citet{belyaev2014}\\
Ca\,II & 69 & 579 & \citet{drawin1969a}, S$_{\mathrm{H}}$=0.1\\
\enddata
\end{deluxetable}

\subsubsection{NLTE abundance results}
We determine an Fe NLTE upper limit abundance of $\mbox{[Fe/H]}<-5.2$ for {\thestar} from an upper limit equivalent width (EW) of $<$10\,m{\AA} for the strongest Fe\,I line at 3859\,{\AA}. We also determined a differential abundance correction, NLTE$-$LTE, value determined self-consistently in \texttt{MULTI2.3}. At this [Fe/H], the correction is +0.4\,dex which we then also apply to the LTE Fe abundance obtained from the composite spectrum using MOOG, leading to $\mbox{[Fe/H]}<-5.4$. Repeating this for our final upper limit of $\mbox{[Fe/H]}<-6.3$ (for which the correction does not significantly increase), shows that the final NLTE upper limit is $\mbox{[Fe/H]}<-5.9$. We note that the true iron abundance of \thestar\ is in all likelihood even lower which would naturally further increase the NLTE effect and associated  correction. By not guessing such a value, we note that we simply produce a more conservative upper limit. However, our calculations do robustly confirm that \thestar\ has an iron abundance of $\mbox{[Fe/H]}<-5.9$. 

We determine NLTE abundances for the $\alpha-$elements, $\mbox{[Na/H]}=-3.92$, $\mbox{[Mg/H]}=-2.82$, $\mbox{[Al/H]}=-3.56$, $\mbox{[Si/H]}=-3.02$ and $\mbox{[Ca/H]}=-5.72$ from 2 Na\,I, 8 Mg\,I, 2 Al\,I, 1 Si\,I and 1 Ca\,II lines, respectively. NLTE effects, determined self-consistently in MULTI, are generally small, to within 0.1\,dex for Na\,I, 0.2\,dex for Mg\,I and 0.05\,dex for Ca\,II. They are slightly higher for Al\,I and Si\,I up to 0.7\,dex and 0.5\,dex, respectively.
For comparison, we also determine NLTE abundances for Na, Mg, Al, Ca and Si for HE~1327$-$2326 \citep{frebeletal05}. NLTE abundances for both {\thestar} and HE~1327$-$2326 are found in Table~\ref{Tab:abund}.

\subsection{The abundance pattern}

\begin{figure}[!t]
 \begin{center}
\includegraphics[clip=true,width=8.cm,viewport=27 80 455 378]{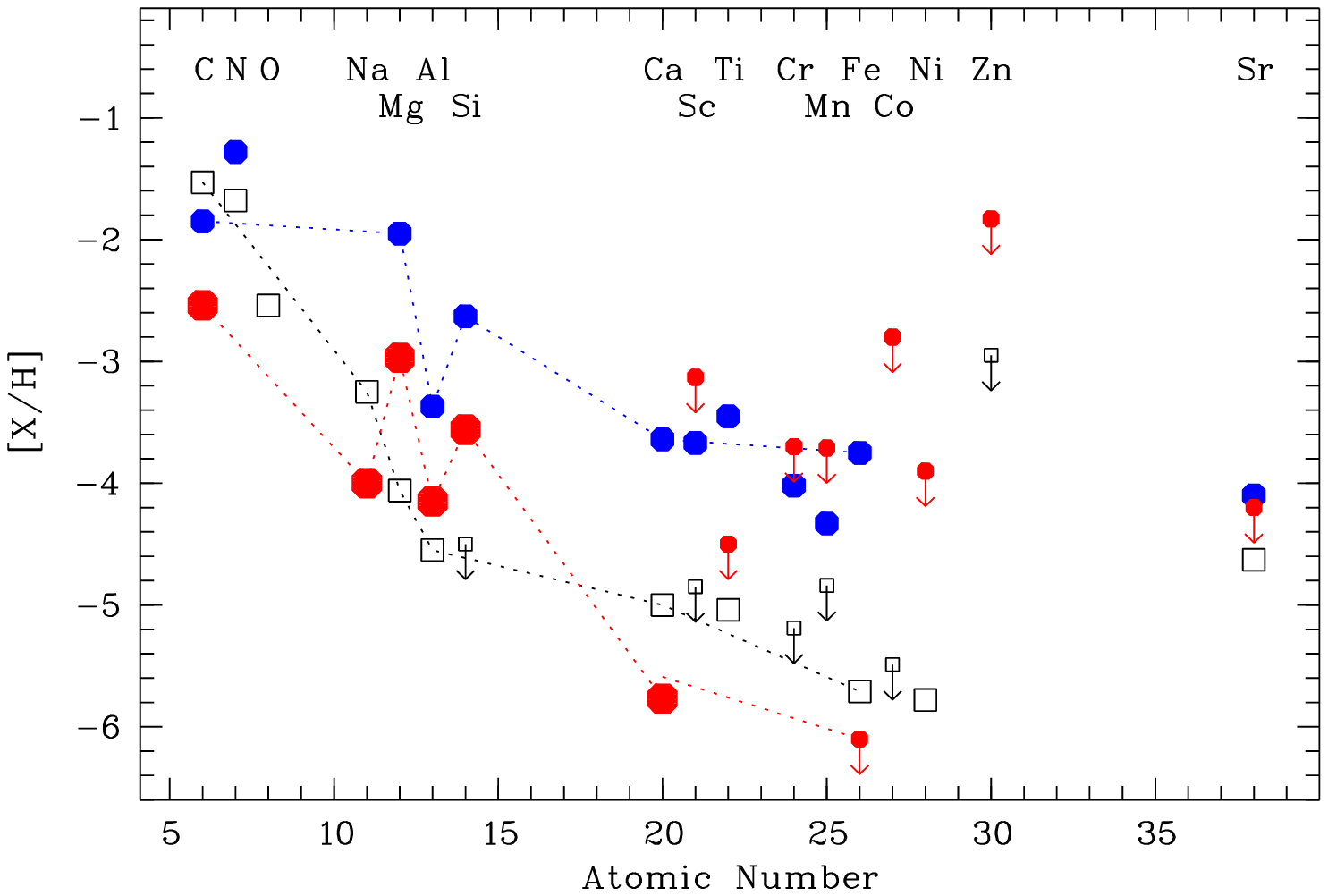}\\
\includegraphics[clip=true,width=8.5cm]{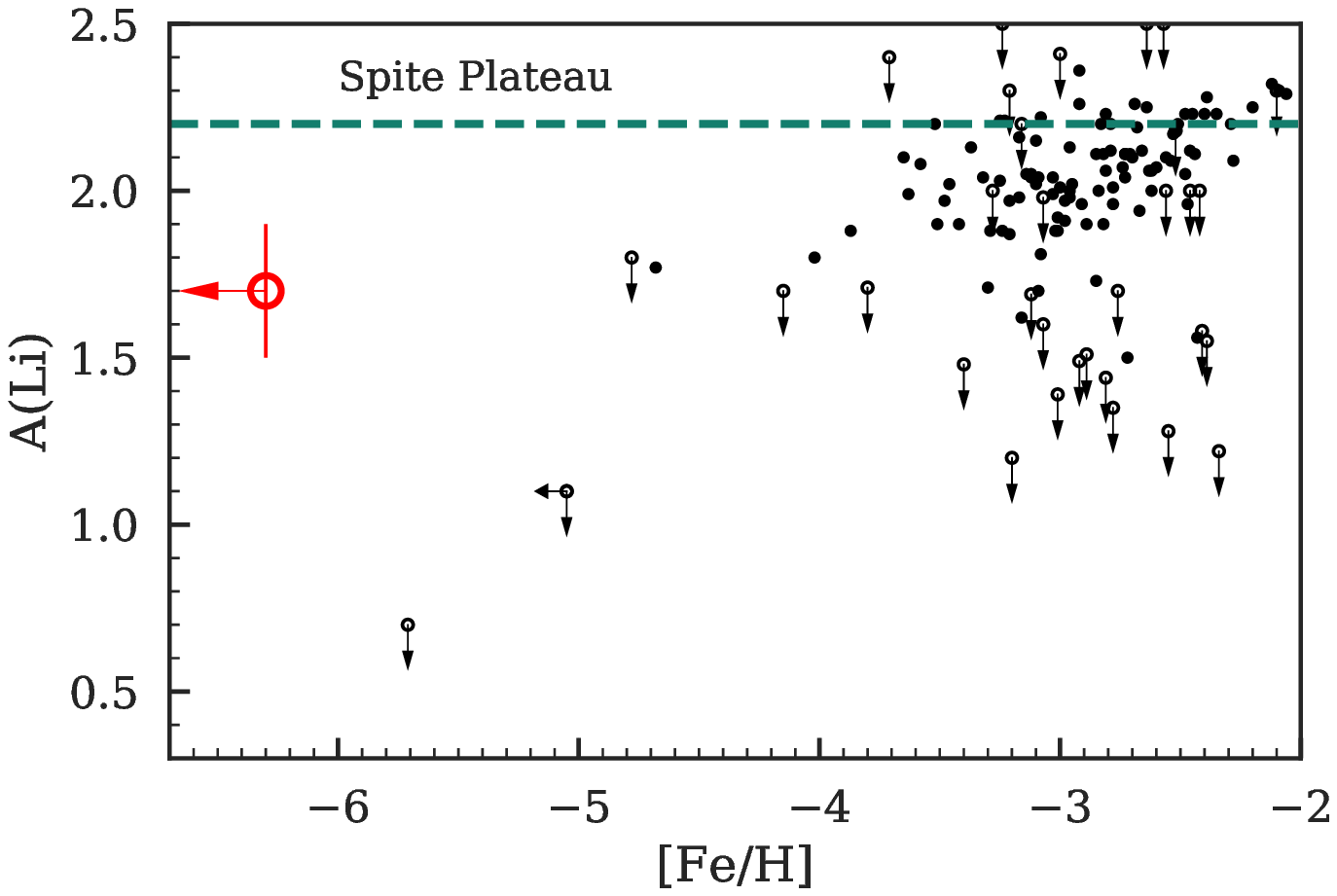}
\figcaption{\label{abund_plot} Top: Abundance patterns of J0023+0307 (red circles), HE~1327$-$2326 (black squares), and CS~29498-043 (blue circles). The same abundances measured in all stars are connected by dotted lines to indicate the similarity of the elemental patterns. Upper limits are denoted with arrows.
Bottom: Li abundances vs [Fe/H]. Comparison halo stars are compiled from JINAbase \citep{jinabase}: \citealt{aoki2008,aoki09b,aoki12,behara10,bonifacio12,Bonifacio15,caffau11,frebel08,fulbright00,hansen15,LiHN15,lucatello03,melendez10,roederer10,Roederer14b,sivarani06,Smiljanic09}.
Stars with $T_{\rm eff} < 5900$\,K and horizontal branch stars have been excluded. Upper limits are denoted with arrows.
}
\end{center}
\end{figure}

In the following, we focus on discussing [X/H] results because Fe is not detected, and its upper limit is very low. However, {\thestar} does show similarities in its abundances compared to four other stars with $\mbox{[Fe/H]}<-5.0$ (when considering 1D LTE values based on a high-resolution spectrum), SDSS~J1035+0641 \citep{Bonifacio15}, HE~0107$-$5240 \citep{HE0107_Nature}, HE~1327$-$2327 \citep{frebeletal05}, and SMSS~0313$-$6707 \citep{keller14}, whose abundances are characterized by the extremely low Fe abundances paired with high C, N, O values. In addition, there are similarities to CS~29498-043 \citep{aoki02} which shares large Mg and Si overabundances, albeit having a much higher iron abundance of $\mbox{[Fe/H]}=-3.75$. A comparison of the 1D LTE elemental patterns of \thestar, CS~29498-043, and HE~1327$-$2327 is shown in Figure~\ref{abund_plot}.

The abundance of Li (A(Li)$=1.7$) is below the value of the Spite Plateau  (A(Li) $\sim 2.20$, \citealt{sbordone10}) for {\thestar} at $\mbox{[Fe/H]}<-6.3$.  The only other warm star with $\mbox{[Fe/H]}<-5.0$ is HE~1327$-$2326 with A(Li)$<0.60$ \citep{frebel08}, but it is worthwhile to also consider SDSS J102915+172927 with $\mbox{[Fe/H]}=-4.7$ and A(Li)$<1.1$ \citep{caffau11} in this context. Both of these have Li abundances much below the Spite Plateau, and thus very different from what is found for \thestar. This is shown in Figure~\ref{abund_plot}. The meltdown of the Spite Plateau below $\mbox{[Fe/H]}\lesssim-3.5$ continues to be discussed in the literature \citep{LiHN15,hansen15b, Matsuno17,Bonifacio18} while explanations for this behavior are sought. \thestar\ adds to the body of data in an [Fe/H] range that is not well populated to date. It helps to address the important issue of whether or not there is a meltdown at the lowest [Fe/H], or perhaps a more varied behavior with no single, well-defined trend. 

The high C abundance of $\mbox{[C/H]}=-2.5$ ($\mbox{[C/Fe]}>3.0$) of \thestar\ is in line with the increased fraction of carbon-enhanced metal-poor stars with decreasing [Fe/H] \citep{placco14}. This means that all five stars with $\mbox{[Fe/H]}<-5.0$ are all strongly carbon-enhanced, with $\mbox{[C/Fe]}>3$. \citet{spite13} find that carbon enhancement at the lowest [Fe/H] values lies at A(C)$\sim6.5$, or $\mbox{[C/H]}\sim-1.9$. The five stars nicely scatter around this value, with \thestar, SMSS0313$-$6708, HE~0107$-$5240, SDSS~J1035+0641, and HE~1327$-$2326 having A(C) $=5.9$, 6.0, 6.8, 6.9, and 6.9, respectively. In a similar way, \citet{Yoon16} postulated that at the lowest [Fe/H], stars should have A(C)$>6.3$. However, both \thestar and SMSS0313$-$6708 with $\mbox{[Fe/H]}<-6.0$ do not show quite this level of enhancement. 

High carbon abundances have previously been explained with carbon (and also oxygen) being present in large enough amounts in the natal cloud ($\mbox{[C/H]}>-3.5$), as provided by the massive first stars. This would have sufficiently cooled the gas to enable the formation of the first low-mass stars in the universe \citep{brommnature,dtrans}. For {\thestar}, this threshold is met even if any systematic 3D effects on the C abundance were of order $-1$\,dex \citep{collet06, gallagher17}.

\thestar\ is the second-most calcium-poor star after SM0313+6708. It also has the highest [Mg/H] abundances of all stars with $\mbox{[Fe/H]}<-4.5$, and the second-highest [Mg/Fe] ratio of all metal-poor stars, after SM0313+6708. These effects remain when considering our NLTE abundances. Overall, \thestar\  shows a strong odd-even effect in the Na through Si abundances. There are also differences between the hydrostatic and explosive $\alpha$-elements: Mg and Si are significantly enhanced in \thestar\ ($\mbox{[Mg/H]}\sim-3.0$, $\mbox{[Si/H]}\sim-3.5$), whereas Ca and Ti abundances are much lower ($\mbox{[Ca/H]}\sim-5.8$, $\mbox{[Ti/H]}<-4.5$).

Such strong odd-even abundance ratios are often associated with pair-instability supernovae \citep[PISN, e.g., ][]{Heger02,aoki14}. However, the lack of significant amounts of Ca and Fe in {\thestar} suggests a PISN could not have been responsible for the observed elements. 
For completeness, we also note that overall, the large C, Mg and Si abundances in combination with the non-detection of Fe clearly point to \thestar\ not being a low-mass ($<1$\,M$_{\odot}$) Population\,III star.

High abundances of Mg and Si have also been found for CS 29498-043 \citep{aoki02} and  J2217+2104 \citep{aoki18} (and also CS 22949-037 but to a lesser extent, \citealt{aoki02}), with a very similar $\alpha$-element pattern to \thestar. Also, both CS 29498-043 and \thestar\ are C-enhanced and share a high Al abundance (see Figure~\ref{abund_plot}). HE~1327$-$2326 and SM0313$-$6708 also display similarly large Mg abundances ($\mbox{[Mg/H]}\sim-4$), and much lower Ca ($\mbox{[Ca/H]}\sim-5$ and $-7$, respectively), all while being C-enhanced. 

\citet{aoki18} find that a supernova with a progenitor mass of 25\,M$_{\odot}$ best fits the abundance patterns of stars with high Mg and Si abundances, especially J2217+2104. This is, however, not different from any progenitor masses thought to explain the patterns of regular metal-poor stars. Accordingly, they suggest that the unusual overabundances are not related to the mass but to other effects in the progenitors such as rotation, mixing, and/or fallback. Alternatively, material released through a stellar wind coming off a rotating massive first star might also explain some of these features \citep{meynet06,Choplin18}. A nitrogen abundance constraint is not possible with our data, but high N could suggest rotation played a significant role.

\begin{figure}[!h]
\includegraphics[width=8.5cm]{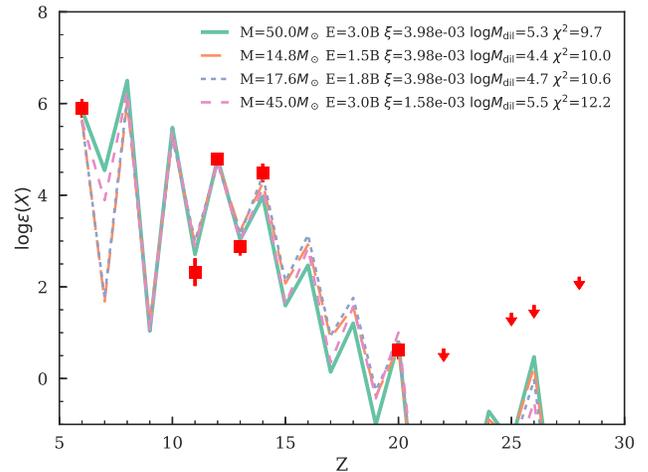}
\caption{\label{fig:fits} The four supernova yield models from \citealt{heger10} (colored lines) that best match the NLTE abundance pattern (red squares).
Model parameters are listed.}
\end{figure}

Some type of massive progenitor, possibly rotating given the large C abundance, must thus have enriched the natal gas cloud of \thestar, either through its supernova yields, or through a stellar wind with the possibility of added supernova yields upon explosion. While the total metallicity of \thestar\ is of course rather high (around $\mbox{[M/H]}\sim-2$ when considering that O and N are likely enhanced at a similar level as C), the low Fe and Ca abundances do clearly point to a single Population\,III star progenitor, as Population\,II supernovae will rapidly erase any low-Fe signature \citep[e.g.,][]{Ji15}. 

\subsection{Fitting Pop\,III supernova yields to the abundance pattern}

We fit Population\,III supernova yields from the non-rotating models of \citet{heger10} to the abundance pattern of \thestar.
We use NLTE abundances, as they are closest to absolute abundances, which are required for a comparison with the yield predictions.

\begin{figure*}
\includegraphics[width=17cm]{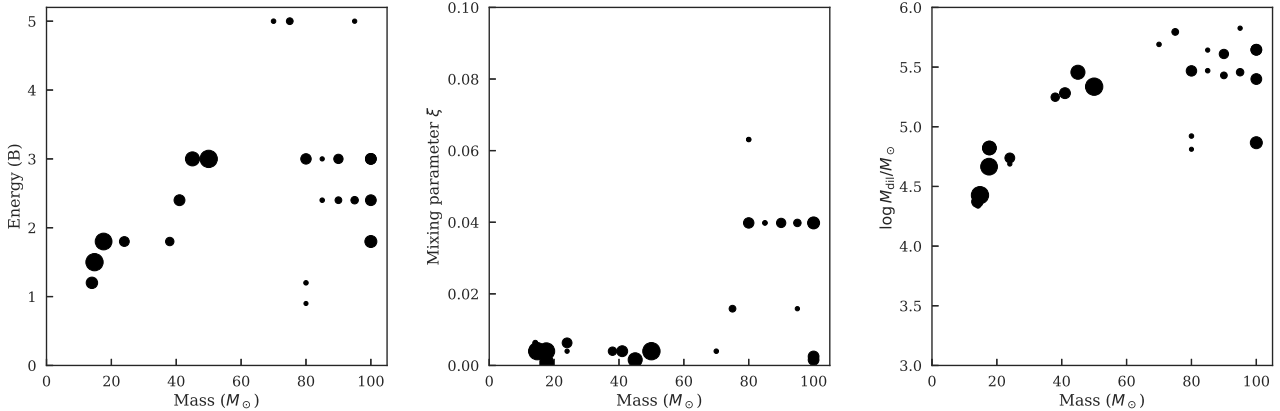}
\caption{\label{fig:allfits} Supernova model parameters of all fits within $2\sigma$ of the data (see text for details). Larger points indicate better fitting model, to draw the eye. The progenitor masses are not constrained, but the energies are moderate ($1-3$\,B), the mixing is low, and the dilution masses range from $10^{4.5-6.0} M_\odot$. Degeneracies exist between progenitor mass and the other parameters.}
\end{figure*}

To find the best-fit models, we determine the best scaling factor in $\log \epsilon$(X) for all 16800 models from \citet{heger10}\footnote{Table \texttt{znuc.S4.star.el.y} from \url{www.2sn.org/starfit}; note that the Starfit algorithm differs from ours in its treatment of upper limits} to minimize $\chi^2$ for our measured element abundances (C, Na, Mg, Al, Si, Ca), i.e.,
\begin{equation}
\chi^2(M,E,\xi,\delta) = \sum_X \left(\frac{\log\epsilon(X) - \log\mu_{M,E,\xi}(X) + \delta}{\sigma_X}\right)^2
\end{equation}
where $\mu_{M,E,\xi}(X)$ is the model abundance of element $X$ in units of $M_\odot/\text{amu}$, with progenitor mass $M$, energy $E$, and mixing parameter $\xi$; and $\sigma_X$ is the abundance uncertainty.
Analytically solving for the minimum gives
\begin{equation}
\delta = -\frac{\sum_X \frac{\log\epsilon(X) - \log\mu_{M,E,\xi}(X)}{\sigma_X^2}}{\sum_X \frac{1}{\sigma_X^2}}
\end{equation}
Note that $\delta < 0$.
Since $\log\epsilon(X) = \log(N_X/N_H) - 12$, the scaling factor $\delta$ can also be converted to a dilution mass $\log M_{\rm dil}$ of hydrogen implied by the model:
\begin{equation}
\log M_{\rm dil} = 12 + \delta
\end{equation}
where the unit of $M_{\rm dil}$ is $M_\odot$, and we use the fact that the atomic mass of hydrogen is 1 amu.
It is unphysical for this dilution mass to be less than the supernova ejecta mass, so we reject all models with $\log M_{\rm dil} < 2$, as well as all models conflicting with our upper limits (Ti, Mn, Fe, Ni).

The best four models are shown in Figure~\ref{fig:fits}.
Qualitatively, the \citet{heger10} models can reasonably reproduce the abundances of {\thestar}, fitting the high carbon, the strong odd-even ratios from Na to Si, the very low Ca, and the low upper limits of heavier elements like Fe.
Several progenitor masses appear to fit almost equally well, but the energies are moderate (1-3B) and mixing parameters $\xi$ are low.
{\thestar} thus appears to be well-described by a fallback-with-little-mixing supernova.
Note that we could not measure N, but it is clear the supernova yields vary greatly in their predicted N abundances, so even a loose constraint on N in the future would be very helpful \citep[also see][]{placco15}.
The N abundance would also useful for determining whether the progenitor star was rotating, as the rotation should greatly enhance N \citep{meynet06,Choplin18}.

The four best-fitting models do not fully represent the complete set of well-fitting models, nor all the parameter degeneracies \citep[e.g.,][]{placco15,Nordlander17}.
Thus, in Figure~\ref{fig:allfits}, we show the best-fit parameters of all models within $\Delta \chi^2 < 9.7$ of the lowest $\chi^2$.
This corresponds to a confidence level of 95\% (or $2 \sigma$), assuming four parameters are fit ($M$, $E$, $\xi$, and $M_{\rm dil}$).
To draw the eye, the size of the point is larger for models that fit better (i.e., lower $\chi^2$).
It is immediately clear that the progenitor mass is not well-constrained, as essentially all masses from $10-100 M_\odot$ have a model consistent within $2 \sigma$ of the measured abundances.
However, the other parameters are better constrained: energies are mostly in the range $1-3 \times 10^{51}$\,erg, the mixing parameters are very low, and the dilution masses are in the range ${\sim}10^{4.5-5.5} M_\odot$.
There are also correlations between mass, explosion energy, and dilution mass.
Since the explosion energy and mixing are not true free parameters but rather an expression of uncertain supernova explosion physics, more realistic 3D simulations \citep[e.g.,][]{Chan18} may eventually be able to break these degeneracies and constrain the progenitor masses.

The dilution masses inferred here strongly suggest that {\thestar} is a second-generation star formed by recollapse in a Population\,III minihalo \citep{Ritter12,cooke14,Ji15}.
In this model, moderate energy Pop\,III supernovae occurring in $10^6 M_\odot$ dark matter minihalos do not evacuate the minihalo of gas, but instead some fraction of the ejected metals recollapse into the same minihalo halo, with typical effective dilution masses ${\sim}10^6 M_\odot$.
The most likely alternate scenarios for second-generation star formation are either external pollution of another minihalo \citep[e.g.,][]{Smith15, Griffen18} or delayed second-generation star formation in an atomic cooling halo \citep[e.g.,][]{Wise08,Greif10}. In either of these cases, we should expect the dilution masses to be much larger, $>10^7 M_\odot$, although for the atomic cooling halo multiple SNe will have contributed metals to the nascent galaxy.
We also note that our somewhat uncertain Na detection has a significant effect on the preferred progenitor masses. Removing the Na constraint causes the fit to prefer lower mass (10-20 $M_\odot$) progenitors, while including Na pushes the fit towards more massive progenitors \citep[also see][]{Ishigaki18}.

\section{Kinematic signature}

\begin{figure*}
\includegraphics[width=16cm, viewport=36 3 950 255]{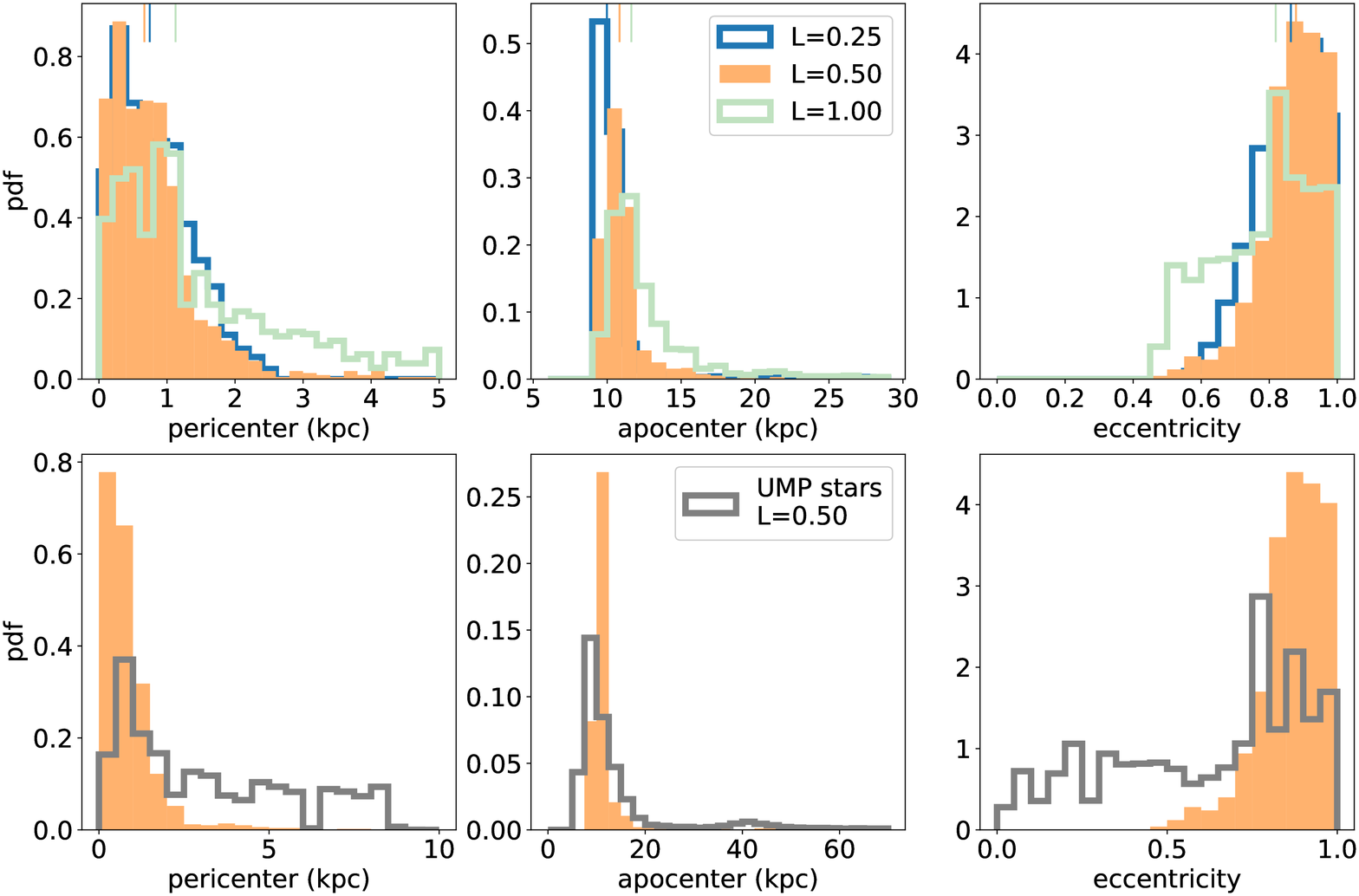}
\caption{Top Row: Posterior distributions for pericenter, apocenter, and eccentricity in {\thestar}. Different colors indicate different assumptions for the distance prior scale $L$. Lines at top indicate median value. 
Bottom Row: Showing comparison to orbits of 25 UMP stars (gray) assuming $L=0.5$ (orange, same data as top panel with different binning).
{\thestar} clearly passes through the bulge at pericenter, and is on a relatively eccentric orbit compared to other UMP stars.
\label{fig:orbit}}
\end{figure*}

We investigate the Galactic orbit of {\thestar} using the parallax and proper motion from Gaia DR2 \citep{GaiaDR2}. Following \citet{BailerJones18} and its extension to proper motions\footnote{\url{https://github.com/agabrown/astrometry-inference-tutorials/blob/master/3d-distance/resources/3D_astrometry_inference.pdf}},
we simultaneously sample the distance and tangential velocity posterior with \texttt{emcee} \citep{emcee}.
For the distance, we use an exponentially decreasing volume density prior \citep{BailerJones18}.
The scale parameter is varied from $L=250$\,pc to $L=1,000$\,pc, where $L=372$\,pc is the value used by \citealt{BailerJones18} based on a magnitude-limited mock survey of a simulated Milky Way.
The total tangential velocity prior is a beta distribution from 0 to 1,800\,km\,s$^{-1}$ with $\alpha=2$ and $\beta=8$. The parameters were chosen to peak at ${\sim}$180\,km\,s$^{-1}$, but changing this prior makes little difference to final results.
Radial velocities are assumed to be normally distributed around $-195$\,km\,s$^{-1}$ with a standard deviation of 5\,km\,s$^{-1}$.
We initialize 400 walkers using the Gaia covariance matrix and assuming $d = 1/\varpi$; burn in 500 steps; then run a chain of length 2,500. Visual examination shows the chains appear converged. 
We then integrate $10^3$ random samples with \texttt{gala} backwards for 3 Gyr in the default \texttt{MilkyWayPotential} \citep{Bovy15,gala} to derive pericenters, apocenters, and eccentricities.
The posterior distributions for the three priors are shown in Figure~\ref{fig:orbit}.

For comparison, we performed the same analysis for other ultra-metal-poor (UMP, $\mbox{[Fe/H]} < -4.0$) stars.
We obtained our list of UMP stars from \citet{jinabase}, supplementing with stars referenced in \citet{Ezzeddine2017} and \citet{Starkenburg18}. After removing two stars with poor parallaxes, our final list contains 26 stars.
Some of these stars are probably binaries, but we do not anticipate radial velocity variations to significantly impact the results for the distribution as the total velocities of most stars are usually relatively high ($>100$ km s$^{-1}$). To somewhat account for this, we include a 5\,km s$^{-1}$ scatter on the radial velocity for all stars. We sampled the kinematic posterior, integrated 1,000 orbits, and show the summed posterior of all stars as a gray histogram in the bottom row of Figure~\ref{fig:orbit}.
Per-star results are in Table~\ref{tbl:umpkin}.

{\thestar}\ is clearly on an eccentric orbit ($e > 0.8$), with a clear pericenter in the bulge ($\lesssim 1$\,kpc).
Its eccentricity is higher and its pericenter smaller than that of a typical $\mbox{[Fe/H]} < -4.0$ star.
Metal-poor stars from the bulge are thought to be older in absolute age \citep{tumlinson10,Howes15,Starkenburg17,ElBadry18}. {\thestar} is thus likely one of the oldest stars known in the Milky Way. Note that of the other two stars with no detected iron, SMSS0313$-$6707 also has a pericenter $<1$\,kpc and is probably also old; but SDSSJ1035$+$0641 does not, with a pericenter at 6.5 kpc.
The highly eccentric radial orbit suggests a possible association with the recently discovered \emph{Gaia} sausage/blob structure \citep{Belokurov18, Koppelman18}, although such orbits could also just reflect typical virialized halo star orbits.
We caution that our simple orbit integrations do not account for effects like the Galactic bar, which can significantly influence halo star orbits \citep[e.g.,][]{PriceWhelan16}.

\begin{deluxetable*}{lllrrrrrrr}
\tablewidth{0pt}
\tablecaption{UMP Star Kinematic Data\label{tbl:umpkin}}
\tablehead{
\colhead{Star} & \colhead{RA} & \colhead{Dec} & \colhead{$v_{\rm hel}$} & \colhead{Distance} & 
\colhead{$v_\alpha$} & \colhead{$v_\delta$} & \colhead{Pericenter} & \colhead{Apocenter} & \colhead{Eccentricity} \\
\colhead{} & \colhead{h:m:s} & \colhead{d:m:s} & \colhead{(km s$^{-1}$)} & \colhead{(kpc)} & 
\colhead{(km s$^{-1}$)} & \colhead{(km s$^{-1}$)} & \colhead{(kpc)} & \colhead{(kpc)} & \colhead{}
}
\startdata
SDSSJ0023$+$0307 & 00:23:14.0 & $+$03:07:58.1 & $-$195 & $2.80^{+0.86}_{-0.65}$ & $49.7^{+15.4}_{-11.7}$ & $-184.8^{+42.4}_{-57.4}$ & $0.67^{+0.61}_{-0.45}$ & $10.84^{+1.24}_{-0.91}$ & $0.88^{+0.08}_{-0.09}$ \\
\hline
HE~0233$-$0343      & 02:36:29.8 & $-$03:30:06.0 &  $+$64 & $1.27^{+0.10}_{-0.09}$ & $301.9^{+23.5}_{-20.2}$ & $-64.0^{+4.2}_{-4.8}$ & $0.30^{+0.40}_{-0.20}$ & $13.40^{+4.31}_{-0.76}$ & $0.95^{+0.03}_{-0.05}$ \\
CS22963$-$004      & 02:56:46.6 & $-$04:51:17.5 & $+$294 & $4.01^{+0.57}_{-0.40}$ & $412.8^{+58.1}_{-41.2}$ & $-50.4^{+5.4}_{-7.5}$ & $1.80^{+1.27}_{-0.81}$ & $72.82^{+59.37}_{-19.11}$ & $0.96^{+0.01}_{-0.01}$ \\
HE~0557$-$4840      & 05:58:39.3 & $-$48:39:56.8 & $+$212 & $9.90^{+1.67}_{-1.28}$ & $33.5^{+6.3}_{-4.9}$ & $34.0^{+6.5}_{-4.9}$ & $2.94^{+0.44}_{-0.29}$ & $14.62^{+1.43}_{-1.09}$ & $0.66^{+0.02}_{-0.02}$ \\
HE~1012$-$1540      & 10:14:53.5 & $-$15:55:53.2 & $+$226 & $0.39^{+0.00}_{-0.00}$ & $-191.0^{+2.2}_{-2.2}$ & $52.5^{+0.6}_{-0.6}$ & $0.88^{+0.08}_{-0.07}$ & $13.58^{+0.14}_{-0.15}$ & $0.88^{+0.01}_{-0.01}$ \\
HE~1310$-$0536\tablenotemark{*} & 13:13:31.2 & $-$05:52:12.5 & $+$113 & $8.61^{+185}_{-1.44}$ & $-198.1^{+38.7}_{-63.5}$ & $-66.4^{+13.2}_{-23.8}$ & \nodata & \nodata & \nodata \\
SDSSJ1808$-$5104   & 18:08:20.0 & $-$51:04:37.9 &  $+$16 & $0.60^{+0.01}_{-0.01}$ & $-15.9^{+0.4}_{-0.4}$ & $-35.8^{+0.9}_{-1.0}$ & $5.19^{+0.11}_{-0.11}$ & $7.82^{+0.04}_{-0.03}$ & $0.20^{+0.01}_{-0.01}$ \\
CS22891$-$200      & 19:35:19.1 & $-$61:42:24.4 & $+$138 & $6.05^{+1.05}_{-0.83}$ & $-146.3^{+19.7}_{-23.6}$ & $22.0^{+3.7}_{-3.0}$ & $0.77^{+0.32}_{-0.54}$ & $10.22^{+1.63}_{-0.49}$ & $0.86^{+0.10}_{-0.05}$ \\
CS22885$-$096      & 20:20:51.2 & $-$39:53:30.2 & $-$249 & $5.31^{+0.59}_{-0.55}$ & $-111.6^{+10.9}_{-12.8}$ & $-173.6^{+17.9}_{-20.1}$ & $4.60^{+0.05}_{-0.21}$ & $7.53^{+0.48}_{-0.36}$ & $0.25^{+0.04}_{-0.03}$ \\
CS22950$-$046      & 20:21:28.4 & $-$13:16:33.6 & $+$107 & $8.04^{+1.35}_{-1.07}$ & $57.9^{+10.1}_{-8.5}$ & $-70.4^{+9.1}_{-11.3}$ & $2.82^{+0.74}_{-0.77}$ & $10.96^{+1.45}_{-0.97}$ & $0.59^{+0.13}_{-0.11}$ \\
CS30336$-$049      & 20:45:23.5 & $-$28:42:35.9 & $-$237 & $9.05^{+1.37}_{-1.16}$ & $-73.8^{+9.1}_{-10.7}$ & $-348.6^{+43.8}_{-53.2}$ & $2.91^{+0.38}_{-0.15}$ & $9.62^{+4.02}_{-2.19}$ & $0.55^{+0.10}_{-0.15}$ \\
HE~2239$-$5019      & 22:42:26.8 & $-$50:04:00.9 & $+$370 & $3.70^{+0.69}_{-0.57}$ & $135.6^{+25.3}_{-20.8}$ & $-415.8^{+63.6}_{-77.0}$ & $7.03^{+0.02}_{-0.07}$ & $34.98^{+30.02}_{-9.89}$ & $0.66^{+0.14}_{-0.10}$ \\
CS22949$-$037\tablenotemark{*} & 23:26:29.8 & $-$02:39:57.9 & $-$125 & $8.96^{+1910}_{-1.75}$ & $69.6^{+5366}_{-14.6}$ & $-74.2^{+15.0}_{-4957}$ & \nodata & \nodata & \nodata \\
BD$+$44$^{\circ}$ 493        & 02:26:49.7 & $+$44:57:46.5 & $-$151 & $0.21^{+0.00}_{-0.00}$ & $118.0^{+1.6}_{-1.5}$ & $-32.1^{+0.4}_{-0.5}$ & $1.01^{+0.08}_{-0.07}$ & $8.66^{+0.04}_{-0.03}$ & $0.79^{+0.01}_{-0.02}$ \\
CD$-$38$^{\circ}$ 245        & 00:46:36.2 & $-$37:39:33.5 &  $+$46 & $4.16^{+0.64}_{-0.44}$ & $300.1^{+46.4}_{-31.4}$ & $-148.7^{+15.5}_{-22.6}$ & $1.19^{+1.26}_{-0.76}$ & $11.83^{+2.47}_{-0.98}$ & $0.82^{+0.11}_{-0.12}$ \\
HE~0057$-$5959      & 00:59:54.1 & $-$59:43:30.0 & $+$375 & $4.70^{+0.57}_{-0.46}$ & $53.2^{+6.6}_{-5.2}$ & $-234.8^{+23.6}_{-28.1}$ & $8.07^{+0.27}_{-0.24}$ & $16.38^{+1.76}_{-1.23}$ & $0.34^{+0.03}_{-0.02}$ \\
HE~0107$-$5240      & 01:09:29.2 & $-$52:24:34.2 &  $+$44 & $7.55^{+1.24}_{-1.00}$ & $86.6^{+14.0}_{-11.8}$ & $-134.3^{+17.1}_{-21.7}$ & $1.38^{+0.62}_{-0.64}$ & $10.48^{+0.78}_{-0.60}$ & $0.77^{+0.11}_{-0.10}$ \\
HE~1327$-$2326      & 13:30:06.0 & $-$23:41:49.7 & $+$112 & $1.13^{+0.03}_{-0.03}$ & $-280.7^{+6.7}_{-8.0}$ & $243.2^{+6.9}_{-6.0}$ & $5.60^{+0.06}_{-0.06}$ & $41.96^{+4.52}_{-3.38}$ & $0.76^{+0.02}_{-0.02}$ \\
HE~2139$-$5432      & 21:42:42.5 & $-$54:18:43.0 & $+$116 & $9.16^{+1.54}_{-1.23}$ & $107.9^{+18.5}_{-14.9}$ & $-195.4^{+26.5}_{-33.5}$ & $0.63^{+0.59}_{-0.41}$ & $8.11^{+1.95}_{-0.85}$ & $0.86^{+0.09}_{-0.14}$ \\
LAMOSTJ1253$+$0753 & 12:53:46.1 & $+$07:53:43.1 &  $+$78 & $0.71^{+0.02}_{-0.02}$ & $71.2^{+2.2}_{-1.9}$ & $-198.6^{+5.3}_{-6.0}$ & $1.66^{+0.09}_{-0.09}$ & $12.13^{+0.20}_{-0.17}$ & $0.76^{+0.01}_{-0.01}$ \\
SMSSJ0313$-$6708   & 03:13:00.4 & $-$67:08:39.0 & $+$300 & $8.18^{+0.86}_{-0.85}$ & $272.2^{+28.8}_{-28.4}$ & $41.5^{+5.2}_{-4.7}$ & $0.92^{+0.19}_{-0.11}$ & $16.39^{+2.72}_{-1.98}$ & $0.89^{+0.02}_{-0.03}$ \\
SDSSJ1204$+$1201   & 12:04:41.4 & $+$12:01:11.5 &  $+$51 & $3.64^{+0.90}_{-0.64}$ & $6.9^{+2.8}_{-2.2}$ & $-85.1^{+14.3}_{-21.8}$ & $4.39^{+0.42}_{-0.56}$ & $9.84^{+0.58}_{-0.37}$ & $0.38^{+0.08}_{-0.06}$ \\
SDSSJ1313$-$0019   & 13:13:26.9 & $-$00:19:41.5 & $+$267 & $3.08^{+0.83}_{-0.63}$ & $-55.1^{+11.1}_{-15.4}$ & $-95.7^{+19.1}_{-25.5}$ & $6.87^{+0.23}_{-0.18}$ & $9.08^{+0.83}_{-0.62}$ & $0.14^{+0.04}_{-0.04}$ \\
SDSSJ1442$-$0015   & 14:42:56.4 & $-$00:15:42.8 & $+$225 & $3.50^{+1.46}_{-1.04}$ & $-4.4^{+5.1}_{-5.9}$ & $110.4^{+47.9}_{-33.6}$ & $6.68^{+0.19}_{-0.34}$ & $31.93^{+17.38}_{-6.00}$ & $0.65^{+0.12}_{-0.07}$ \\
SDSSJ1029$+$1729   & 10:29:15.2 & $+$17:29:27.9 &  $-$34 & $1.41^{+0.18}_{-0.13}$ & $-72.7^{+7.3}_{-9.3}$ & $-27.0^{+2.5}_{-3.4}$ & $8.03^{+0.23}_{-0.24}$ & $9.03^{+0.14}_{-0.09}$ & $0.06^{+0.02}_{-0.02}$ \\
HE~1424$-$0241      & 14:26:40.3 & $-$02:54:27.5 &  $+$60 & $5.42^{+1.16}_{-0.93}$ & $-97.9^{+16.4}_{-21.2}$ & $-72.8^{+12.8}_{-16.2}$ & $2.51^{+0.51}_{-0.63}$ & $6.79^{+0.19}_{-0.04}$ & $0.46^{+0.12}_{-0.08}$ \\
SDSSJ1035$+$0641   & 10:35:56.1 & $+$06:41:44.0 &  $-$78 & $3.49^{+1.42}_{-1.09}$ & $59.9^{+26.6}_{-19.0}$ & $-47.7^{+14.0}_{-18.4}$ & $6.50^{+0.52}_{-0.65}$ & $19.08^{+4.08}_{-2.35}$ & $0.49^{+0.10}_{-0.08}$ \\
SDSSJ1742$+$2531   & 17:42:59.7 & $+$25:31:35.9 & $-$208 & $3.65^{+1.26}_{-0.81}$ & $-109.5^{+24.3}_{-36.6}$ & $-198.4^{+44.3}_{-72.2}$ & $0.90^{+0.18}_{-0.36}$ & $7.42^{+1.75}_{-0.17}$ & $0.79^{+0.10}_{-0.04}$ \\
Pristine\_221     & 14:47:30.7 & $+$09:47:03.7 & $-$149 & $4.04^{+1.03}_{-0.73}$ & $-149.0^{+27.8}_{-38.6}$ & $-1.0^{+2.1}_{-2.2}$ & $3.46^{+0.35}_{-0.27}$ & $9.54^{+0.40}_{-0.19}$ & $0.47^{+0.04}_{-0.05}$ \\
\enddata
\tablenotetext{*}{Not used due to large uncertainty.}
\tablecomments{Estimate is median of 1,000 samples, uncertainty is 16-84th percentiles. All values assume distance prior with $L=500$pc.}
\end{deluxetable*}

\section{Summary and Conclusions}

We have presented a high-resolution spectrum of the iron-poor main-sequence \thestar. Iron lines were not detected but the $S/N$ of the spectrum is only moderate. Stacking 19 Fe lines into a composite spectrum did not yield a detection either but led to an upper limit of $\mbox{[Fe/H]}<-5.8$. Invoking also that the Ca abundances in the most iron-poor stars always exceed the Fe abundance, we deduce an upper limit of $\mbox{[Fe/H]}<-6.3$. Lines of other elements are detected, i.e., Li, C, Na, Mg, Al, Si and Ca. Lithium has been challenging to determine given the somewhat distorted line. We thus cannot draw any firm conclusions on the Li abundance, beyond suggesting that the two warm stars with $\mbox{[Fe/H]}<-5.0$ both have lower than Spite Plateau values. However, the range covered by those low Li abundances is large ($\sim1$\,dex) adding to the complex behavior of Li found in the most iron-poor stars. Higher quality data would be able to refine the Li abundance to gain more insight into the evolution of Li in the early universe.

There is a strong odd-even effect in the abundances from Na to Si. \thestar\ has the second-lowest Ca abundance of any know star after SMSS0313$-$6708, by far the highest [Mg/Fe] ratio of any metal-poor stars (together with SMSS0313$-$6708) and also among the highest [Mg/H] of stars with $\mbox{[Fe/H]}<-4.0$. Carbon is detected and strongly C enhanced, in line with values found for the other of the most iron-poor stars. This likely points to fragmentation from early gas cooling by C and O provided by Population\,III first stars.

The overall metallicity of \thestar\ is $\mbox{[M/H]}\sim-2.0$ which is largely driven by the large C (and presumably large N and O) abundance. Nevertheless, the low Fe and Ca abundances clearly point to \thestar\ being a second generation star that formed from gas enriched by just one Population\,III supernova. Fitting the abundance pattern with nucleosynthesis yields of the first stars shows that the mass of the progenitor star is essentially unconstrained given current abundance uncertainties, but intermediate explosion energies of 1-$3\times 10^{51}$\,erg and a low mixing parameter are likely.
The inferred dilution masses from these fits are $10^{4.5-5.5} M_\odot$ of hydrogen, as is expected if \thestar\ formed in a recollapsed minihalo.
\thestar\ has a rather eccentric orbit ($e>0.8$) which is among the more eccentric values of 25 stars with $\mbox{[Fe/H]}<-4.0$. The pericenter takes the star right through the Galactic bulge, suggesting {\thestar} is one of the oldest stars.

Additional data should be sought to attempt a detection of Fe lines and to produce even tighter upper limits on various so far undetected elemental abundances. This will be important for further constraining the nature of the progenitor of \thestar. More data would also assist in confirming the Li abundance to firmly establish the range of observed (i.e., physically possible) Li abundances at $\mbox{[Fe/H]}<-5.0$.
Thus far, no radial velocity variations have been detected; future measurements would also help to further constrain the nature and formation scenario of \thestar.


\acknowledgements 
We thank Conrad Chan for assistance with the supernova yield models.
A.F. is partially supported by NSF-CAREER grant AST-1255160 and NSF grant 1716251.  
A.P.J. is supported by NASA through Hubble Fellowship grant HST-HF2-51393.001 awarded by the Space Telescope Science Institute, which is operated by the Association of Universities for Research in Astronomy, Inc., for NASA, under contract NAS5-26555.
R.E. acknowledges support from a JINA-CEE fellowship (Joint Institute for Nuclear Astrophysics - Center for the Evolution of the Elements), funded in part by the NSF under Grant No. PHY-1430152 (JINA-CEE).
%
%
This work made use of NASA's Astrophysics Data System Bibliographic Services, and the SIMBAD database, operated at CDS, Strasbourg, France \citep{simbad}.

This work has made use of data from the European Space Agency (ESA) mission
{\it Gaia} (\url{https://www.cosmos.esa.int/gaia}), processed by the {\it Gaia}
Data Processing and Analysis Consortium (DPAC,
\url{https://www.cosmos.esa.int/web/gaia/dpac/consortium}). Funding for the DPAC
has been provided by national institutions, in particular the institutions
participating in the {\it Gaia} Multilateral Agreement.

\facilities{Magellan-Clay (MIKE, \citealt{mike})}

\software{MOOG~\citep{Sneden73,Sobeck11}, MIKE Carnegie Python Pipeline~\citep{Kelson03},
MULTI~\citep{Carlsson1986,carlsson1992},
MARCS~\citep{gustafsson1975,gustafsson2008},
IRAF~\citep{irafa, irafb}, NumPy~\citep{numpy}, SciPy~\citep{scipy}, Matplotlib~\citep{matplotlib}, Astropy~\citep{astropy,astropy2},
emcee~\citep{emcee,emceeASCL},
gala~\citep{gala,galazenodo},}

\end{document}